\documentclass[useAMS,usenatbib]{mn2e}
\usepackage{amssymb,amsmath,epsfig,times, natbib,color}
\pdfoutput=1

\title[Coma outskirts]{A complete view of the outskirts of the Coma cluster}
\author[M. S. Mirakhor et al] {\parbox[]{6.5in}{{
      M. S. Mirakhor$^1$\thanks{Email: 
    msm0033@uah.edu} and S. A. Walker$^1$
    }\\
     \footnotesize
     $^1$Department of Physics and Astronomy, University of Alabama in Huntsville, Huntsville, AL 35899, USA \\
 }}

\date{}

\begin{document}

\maketitle

\begin{abstract}
We present a new extended \textit{XMM-Newton} mosaic of the nearby Coma cluster, which covers the cluster out to the virial radius with nearly complete azimuthal coverage. This large mosaic is combined with the \textit{Planck} Sunyaev Zel'dovich effect observations to recover the thermodynamic properties of the intracluster medium in an azimuthally averaged profile and 36 angular sectors, producing the highest spatial resolution view of the thermodynamics of the outskirts of a galaxy cluster in its entirety. Beyond $r_{500}$, our clumping corrected entropy measurements along the less disturbed directions are statistically consistent with the power-law entropy profile predicted by non-radiative simulations, and the gas mass fraction agrees with the mean cosmic baryon fraction. However, there is a clear entropy deficit in the outskirts to the southwest, coinciding with where Coma connects to a cosmic web filament which joins it to Abell 1367. The low entropy to the southwest extends from 0.5-1.0$r_{200}$, and is consistent with what is expected from simulations of a filamentary gas stream penetrating into the cluster as it continues to accrete matter from the cosmic web. We also find that the radial profiles of the recovered quantities become increasingly asymmetric in the outskirts, particularly along the more disturbed directions, consistent with the predictions of cosmological simulations.

\end{abstract}

\begin{keywords}
galaxies: clusters: intracluster medium - intergalactic medium
- X-rays: galaxies: clusters
\end{keywords}

\section{Introduction}
Accurate measurements of the intracluster medium (ICM) of galaxy
clusters to the virial radius are important for many reasons. The virial
radius represents the boundary within which the cluster ICM is expected
to be virialised, and outside of which matter is still accreting onto
the cluster as it continues to form. Studying cluster outskirts allows
the cluster formation process to be understood, thus constraining
simulations of cluster formation and models for the baryon fractions of
clusters. We can also investigate where the assumption of hydrostatic
equilibrium breaks down, which is important for calculating the masses
of galaxy clusters, and using clusters as cosmological probes. For a review see \citet{Walker2019}.

Due to the very low X-ray surface brightness in cluster outskirts, accurate measurements of the ICM around the virial radius are highly challenging. Systematic uncertainties in the modelling of the \textit{XMM-Newton} background mean that finding gas temperatures through spectral fitting outside $r_{500}$ (which is around 65 percent of the virial radius, $r_{200}$) is generally impossible with \textit{XMM-Newton}. One new approach, which gets around these limitations, is to combine X-ray surface brightness measurements with \textit{Planck} Sunyaev Zel’dovich effect (SZ) measurements \citep{Tchernin2016, Eckert2017, ghirardini2018xmm, Ghirardini2019}. For plasma with temperatures above 1.5 keV, the X-ray surface brightness in the soft band 0.7-1.2 keV can be trivially converted into gas density, as it has very little dependence on the gas temperature. These gas densities, $n_{e}$, can then be combined with the pressure measurements, $P$, from \textit{Planck} to obtain the gas temperature, $kT=P/n_{e}$, and the gas
entropy, $K=kT/n_{e}^{2/3}=P/n_{e}^{5/3}$.

Simulations of galaxy cluster formation (e.g. \citealt{Roncarelli2006}, \citealt{vazza2011scatter}, \citealt{avestruz2014testing}, \citealt{lau2015mass}) predict that clusters become increasingly asymmetric in both density and temperature in the outskirts, where the cluster joins onto the surrounding cosmic web. Such simulations predict azimuthal variations in the region between $r_{500}$ and $r_{200}$ due to the asymmetric nature of accretion onto galaxies clusters from the cosmic web, and these can only be resolved by dividing the cluster into small enough sectors. Simulations (e.g. \citealt{Gaspari2013}) indicate that the peak of the power spectrum of gas motions in galaxy clusters should lie in the range 250-500kpc, so it is important to be able to study the ICM properties on these scales. The simulations of \cite{zinger2016role, Zinger2018} have found that the gas streams from the filaments of the cosmic web can penetrate deep into unrelaxed clusters, transporting low entropy gas and large amounts of momentum and energy into their central regions.

At present, unravelling and mapping this fine structure in the outskirts of clusters has proven difficult for a number of reasons. As a result, our measurements of cluster outskirts are typically limited to azimuthally averaged profiles, which are clearly not telling us the full story of what is happening in the outskirts. 

Direct X-ray observations of gas temperatures
with \textit{Suzaku}, which had a much lower and more stable background than \textit{XMM-Newton} (\textit{Suzaku}'s particle background was a quarter of that of \textit{XMM-Newton}'s due to \textit{Suzaku}'s low Earth orbit) were limited in their spatial resolution due to \textit{Suzaku}'s large point spread function (PSF) of 2 arcmin, and limited in their azimuthal coverage by Suzaku's small field of view of 17.8 arcmin $\times$ 17.8 arcmin (see \citealt{Walker2019} for a review). Observing strategies with Suzaku could be divided into two groups. The first focused on high azimuthal coverage of higher redshift clusters with low spatial resolution (e.g. \citealt{kawaharada2010suzaku}, \citealt{Walker2012_A2029}a, \citealt{Walker2012_PKS0745}b, \citealt{Sato2012}, \citealt{ichikawa2013suzaku}). The second focused on getting around Suzaku's large PSF by looking at nearby clusters to achieve high spatial resolution (\citealt{Simionescu2011}, \citealt{Urban2014}), Centaurus (\citealt{Walker2013_Centaurus}), Coma (\citealt{simionescu2013thermodynamics}) and Virgo (\citealt{Simionescu17}). However the small field of view of \textit{Suzaku} meant that the coverage of these nearby clusters was limited to strips covering at most 30 percent of the virial radius. 

The method of combining XMM surface brightness with \textit{Planck} pressure measurements is also limited by the large PSF of \textit{Planck} ($\sim$10 arcmin FWHM). To make progress in understanding the detailed physics in the outskirts of clusters, we therefore need to study low redshift clusters, whose large angular extents allow us to circumvent the large PSF of \textit{Planck}. Thanks to XMM's larger field of view an approximately circular field with a diameter of 30 arcmin), and the fact that we
only require surface brightness measurements to combine with the existing \textit{Planck} pressures rather than detailed X-ray spectroscopy, mapping campaigns of nearby clusters with full coverage out to their virial radii can provide us with the best ever view of the outskirts of galaxy clusters. 

As described by the Planck collaboration itself, the Coma cluster is `the most spectacular SZ source in the Planck sky' (\citealt{ade2013planck}).
 As one of the nearest massive clusters in the universe ($M_{500} \approx 6.0 \times 10^{14}$ M$_{\odot}$ at redshift 0.023), \textit{Planck} 
was able to detect the SZ emission in Coma with extremely high signal to noise ($>$ 22) over a 5$\times$5 degree area. \textit{Planck} was able to measure the SZ signal in Coma out to colossal radii, out to 2 times the virial radius (2$r_{200}$).
   
At the low redshift of Coma, the \textit{Planck} beam size of 10 arcmin (FWHM) corresponds to a physical size of 280 kpc, which is sufficient resolution to test the predictions of numerical simulations for the asymmetric distribution of temperature and entropy in cluster outskirts. \citet{malavasi2020like} have used Sloan Digital Sky Survey (SDSS) to show that the Coma cluster is a highly connected structure, with three main connections to cosmic web filaments to the northeast, north and south west. This makes Coma an excellent target to study the azimuthal variation of the ICM's thermodynamic properties, and how these relate to Coma's larger scale environment.



Here we present our new extended \textit{XMM-Newton} mosaic of the Coma cluster, which covers the cluster out to $r_{200}$ with nearly complete azimuthal coverage (the XMM mosaic covers 90 percent of the cluster inside $r_{200}$). By combining the X-ray surface brightness information from this mosaic with the \textit{Planck} SZ data for Coma, we are able to produce the highest spatial resolution view of the thermodynamics of the outskirts of a galaxy cluster ever achieved.

\citet{ade2013planck} estimated the $Y_X=M_{\rm{gas}} \times kT$ parameter of the Coma cluster iteratively using the $Y_X-M_{500}$ scaling relation calibrated from hydrostatic mass estimates in a nearby cluster system \citep{Arnaud2010}, and found that the scale radius of Coma at a density 500 times the critical density of the universe to be $r_{500}=47$ arcmin. We adopt this value throughout the paper, and this implies the scale radius $r_{200}\approx 70$ arcmin, using the relation $r_{500} = 0.65 r_{200}$ obtained assuming an NFW profile with a concentration parameter of 4. These are the same values used in the study of the Suzaku observations in \citet{simionescu2013thermodynamics}.

The paper is structured as follows: in Section \ref{sec: data} we describe the X-ray and SZ effect data used in this work; the analysis procedures for our data are presented in Section \ref{sec: analysis}; in Section \ref{sec: joint xray and sz} we show the reconstructed profiles of the thermodynamic properties of the ICM; we present the analysis in angular sectors in Section \ref{sec: sector analysis}; our findings are discussed in Section \ref{sec: discussion}; and our conclusions are presented in Section \ref{sec: conclusions}. Throughout this paper, we adopt a $\Lambda$ CDM cosmology with $\Omega_{\rm{m}}=0.3$, $\Omega_{\rm{\Lambda}}=0.7$, and $H_0=100\,h_{100}$ km s$^{-1}$ Mpc$^{-1}$ with $h_{100}=0.7$. At the redshift of Coma, 1 arcmin corresponds to 28 kpc.

\begin{figure*}
\begin{center}

\includegraphics[width=1.0\textwidth]{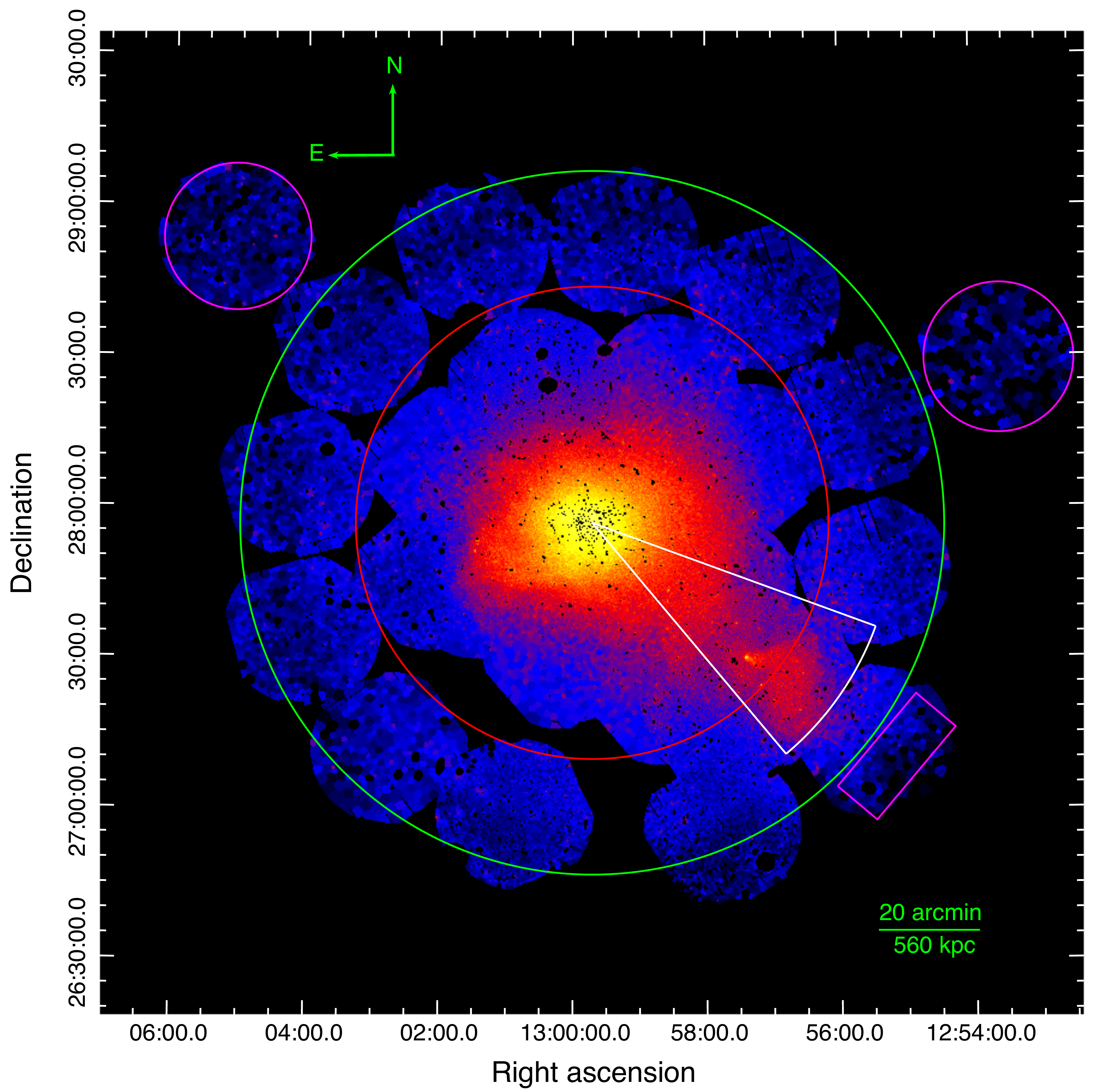}

\end{center}
\vspace{-0.5cm}
\caption{\textit{XMM-Newton} mosaicked and Voronoi tessellated image of the Coma cluster in the 0.7–1.2 keV energy band. North is up and east is to the left. The red and green circles show, respectively, the locations of $r_{500}$ and $r_{200}$. The magenta regions show the regions in which the local background counts were extracted, and the white sector indicates the region which is excluded from the analysis as it contains the infalling group NGC 4839.}
\label{fig: coma_xray}
\end{figure*}

\section{Data}
\label{sec: data}
\subsection{\textbf{\textit{XMM-Newton}}}
A large mosaic of \textit{XMM-Newton} observations was obtained during the AO-18 cycle (PI: S. A. Walker). The new mosaic consists of 11 pointings, with 9 of which covering the outskirts region of the cluster between $r_{500}$ and $r_{200}$, whereas 2 pointings covering the local sky background of the cluster in two different positions. The exposure time ranges between 23 and 35 ks each, for a total exposure time of 300 ks. We combined these observations with the 50 existing \textit{XMM-Newton} observations in the field of the Coma cluster that cover, mainly, the central and intermediate regions of the cluster. These archive observations were performed in the period between 2000 and 2019, and their total observing time is around 2.0 Ms. The details regarding all 61 of the \textit{XMM-Newton} observations used in this work are summarized in Table \ref{tab: xmm_observations}.

\subsection{\textbf{\textit{Planck}}}
The Coma Cluster was also observed by the \textit{Planck} satellite in nine frequency bands covering 30$-$857 GHz. The SZ effect signal detected towards Coma is significant with a signal to noise ratio $>$ 22 \citep{collaboration2011planck}. \textit{Planck} was able to measure the SZ effect signal in Coma out to large radii, out to at least two times the virial radius. The data used in this paper based on the \textit{Planck} nominal survey that carried out between 2009 August 13 and 2010 November 27.

\section{Analysis Procedures}
\label{sec: analysis}

\subsection{\textbf{\textit{XMM-Newton} analysis}}
\label{sec: XMM_analysis}
The X-ray data for all 61 observations in this work were reduced using the \textit{XMM-Newton} Science Analysis System, XMM-SAS v18.0, and the calibration files, following the procedures illustrated in the Extended Source Analysis Software manual \citep[ESAS,][]{snowden2014cookbook}. We initially processed the data using the \textit{epchain} and \textit{emchain} scripts, and then used ESAS tasks \textit{mos-filter} and \textit{pn-filter} to remove soft proton flares and create clean event files. The data were then examined for CCDs in anomalous states, and any affected CCDs were excluded from the analysis. Point sources and extended substructures that contaminated the field of view were masked and removed by running the ESAS tool \textit{cheese}. For all \textit{XMM-Newton} detectors, the required intermediate spectra and response files for the entire region of interest were created using the \textit{mos-spectra} and \textit{pn-spectra} tasks. These files were then used to create the quiescent particle background (QPB) spectra and images by running the ESAS procedures \textit{mos-back} and \textit{pn-back}.

We also filtered the data for any residual soft proton contamination that may have remained after the initial light curve screening. The filtering was performed by fitting the spectral data to a model with components representing emission from the residual soft proton contamination, the target, the cosmic diffuse X-ray background, Gaussian lines, and solar wind charge exchange lines. The derived values of the spectral parameters for the soft proton contamination were then used in the ESAS task \textit{proton} to create images of the soft proton contamination. 

The analysis procedures described above created all of the primary components for a background-subtracted and exposure-corrected image. These components from all observations and three \textit{XMM-Newton} detectors were then merged into a single image, weighting each detector by its relative effective area, and the resulting image adaptively smoothed. Following \citet{ghirardini2018xmm}, the created image was also filtered by the \textit{Chandra} tool \textit{wavdetect} to detect and remove any remaining point sources and extended substructures that missed using the ESAS tool \textit{cheese}. To create an adaptively binned image with binning the counts to at least 20 per energy bin, we applied a Voronoi tessellation algorithm \citep{diehl2006adaptive} on the mosaicked count image.

Fig. \ref{fig: coma_xray} shows the spatial coverage of our \textit{XMM-Newton} mosaicked and Voronoi tessellated image of the Coma cluster in the 0.7$-$1.2 keV energy band. We chose this energy range in order to minimize the high-energy-particle background, which rises significantly at low and high energies. The X-ray structure shows that the ICM gas in Coma extends in the west and southwest directions towards the infalling substructure NGC 4839. 

The imaging-data analysis of galaxy clusters is subject to various sources of systematic uncertainty. The choice of the local background region, particularly at the cluster outskirts, is the major source of systematic uncertainty in the X-ray image of a galaxy cluster. For this purpose, we chose three background regions in three different locations outside the $r_{200}$ radius of Coma, as shown in the magenta regions in Fig. \ref{fig: coma_xray}, to extract the local X-ray background counts. The magenta region in the northwest is chosen as the local background for the cluster's regions located in the west and northwest directions, and the one in the southwest direction is chosen as the local background for the regions located in the southwest and south directions. For other regions, the magenta region in the northeast is chosen as the local background. Moreover, we considered a $\pm 5\%$ uncertainty of the background level as an additional uncertainty in the radial profile of the X-ray surface brightness to account for background fluctuations, and is added in quadrature to each annulus in the count rates. All measurements reported in this paper take account of the systematic uncertainties discussed above.   

\citet{Zhuravleva2013} have found that the distribution of gas densities in a given region of the ICM is well described as a log-normal component describing the bulk of the gas, and a high density tail resulting from gas clumping. In order to accurately measure the clumping factor, we therefore need a sufficient number of Voronoi tesselated regions in each region we study to be able resolve the shape of the surface brightness distribution into these two components (the log-normal distribution and the tail). We find that a minimum of 150 independent Voronoi regions is needed to be able to accurately resolve the shape of the surface brightness distribution to calculate the median, and we ensure that this threshold is met for all of the regions we study both here and in our sector analysis in section \ref{sec: sector analysis}.

\begin{figure}
\begin{center}
\includegraphics[width=\columnwidth]{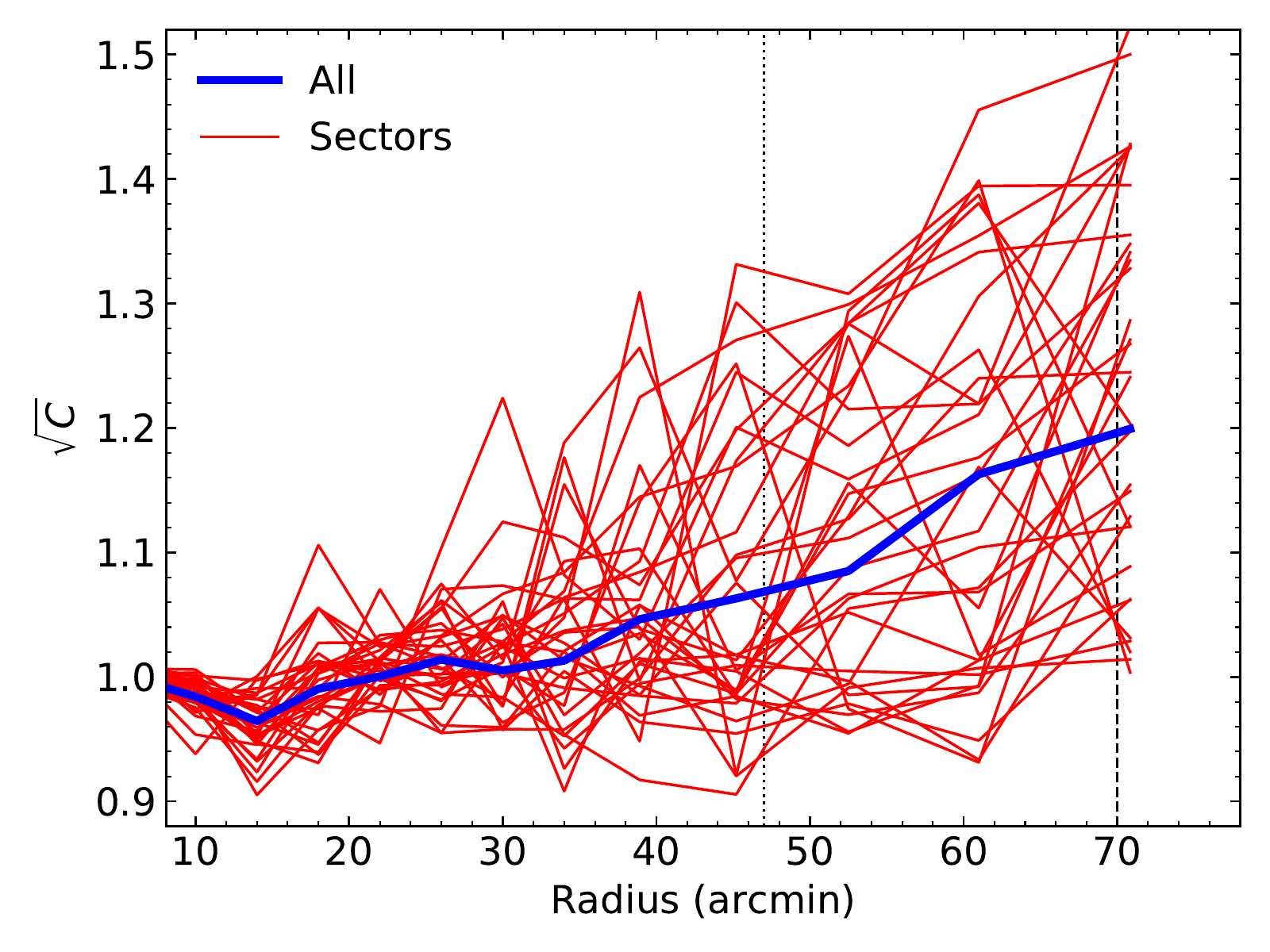}
\end{center}
\vspace{-0.5cm}
\caption{Radial profile of clumping factor for the azimuthally averaged profile (blue) and 33 angular sectors (red) discussed in Section \ref{sec: sector analysis}. The vertical dotted and dashed lines represent the $r_{500}$ and $r_{200}$ radii, respectively.}
\label{fig: clumping}
\end{figure}

To derive surface brightness profile for our mosaicked image of the Coma cluster, we extracted counts in concentric annuli centered at the cluster centre, (RA, Dec.) = (12:59:42.44, +27:56:45.53) This is the same cluster center as used in \citet{simionescu2013thermodynamics}, which we use for consistency and is the location of the X-ray peak. We tested the effect of changing the central coordinate used, and found this to have no significant effect on our results. We excluded the region that contains the NGC 4839 substructure from our analysis (the white sector in Fig. \ref{fig: coma_xray}). When extracting the radial profile of the surface brightness profiles, particularly in the outskirts of clusters, it is important to consider possible biases introduced by gas clumping, which become increasingly important near the virial radius \citep{Walker2019}. We estimated the level of gas clumpiness directly from the X-ray image by dividing the mean surface brightness to the median surface brightness in each annulus region, as illustrated in \citet{eckert2015gas}. The inferred clumping factor $\sqrt{C}$ is shown in Fig. \ref{fig: clumping}. Within $r_{500}$, the estimated values of the clumping factor are low and show a slight trend of increasing with radius. Beyond $r_{500}$, the clumping value increases to around 1.2, consistent with the range of values typically found in the outskirts \citep{eckert2015gas,Walker2019}.  

\begin{figure}
\begin{center}
\vbox{
\includegraphics[width=\columnwidth]{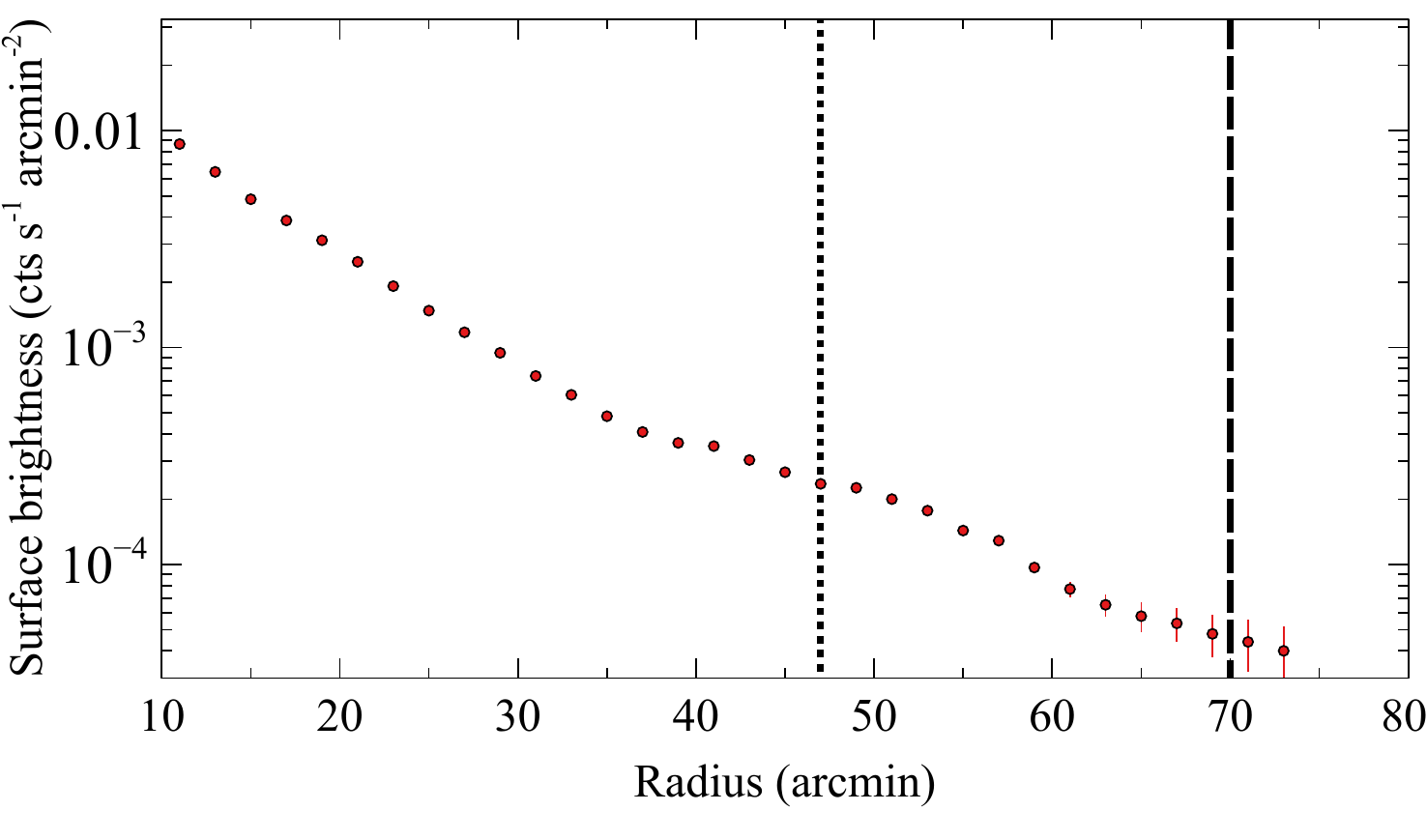}
\includegraphics[width=\columnwidth]{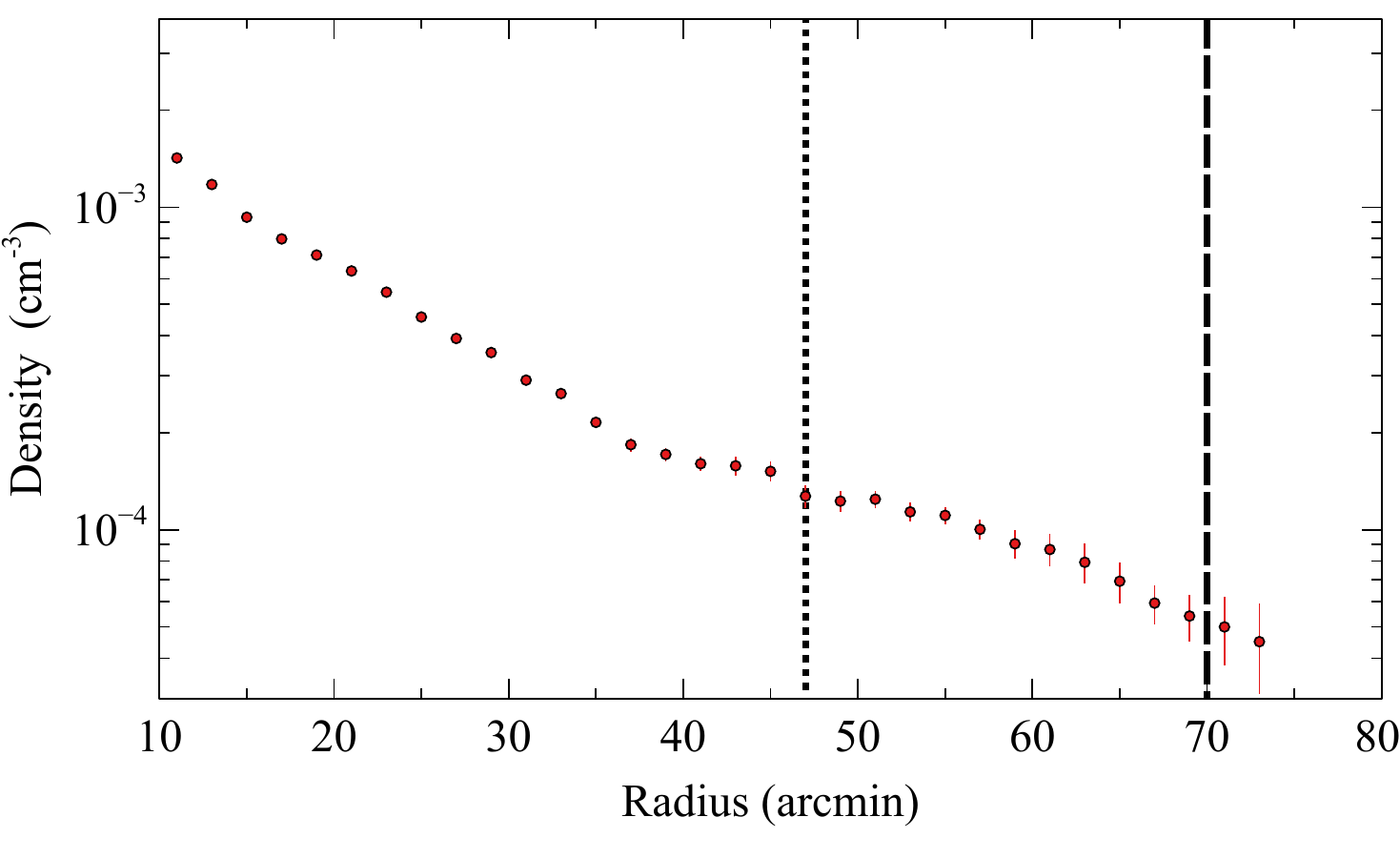}
}
\end{center}
\vspace{-0.5cm}
\caption{\textit{Top}: Azimuthal median of the background-subtracted surface brightness profile for the Coma cluster in the 0.7$-$1.2 energy band. \textit{Bottom}: Density profile obtained from the deprojection of the surface brightness radial profile using the onion peeling technique. The error bars are the 1$\sigma$ percentiles computed using a Monte Carlo technique. The vertical dotted and dashed lines represent the $r_{500}$ and $r_{200}$ radii, respectively.}
\label{fig: sb_ne_all}
\end{figure}

In the upper panel of Fig. \ref{fig: sb_ne_all}, we show the azimuthal median of the background-subtracted surface brightness profile for Coma in the 0.7$-$1.2 keV energy band. The electron density, shown in the lower panel of Fig. \ref{fig: sb_ne_all}, was then recovered from the deprojection of median surface brightness profile using the onion peeling technique \citep{Fabian1981,Kriss1983,ettori2010mass}, and assuming that the ICM plasma is spherical symmetry.

For the electron density, we follow the same approach as \citet{Tchernin2016} and \citet{Ghirardini2019}. We fold an APEC model (absorbed by the galactic column density, $N_{\rm{H}}=9.3\times10^{19}$ cm$^{-2}$ obtained from the Leiden-Argentine-Bonn survey, \citealt{LABsurvey}) through the \textit{XMM-Newton} response in XSPEC to obtain the conversion between the count rate observed by \textit{XMM-Newton} in the 0.7-1.2 keV band and the APEC normalization. As described in \citet{Tchernin2016}, the 0.7-1.2 keV count rate is largely independent of temperature as long as the temperature is above $\sim$1.5 keV. For consistency with \citet{simionescu2013thermodynamics}, we use the same abundance table \citep{Grevesse1998} and the same metal abundance value in the outskirts of 0.3$Z_{\odot}$. The APEC model normalization is related to the gas density as
\begin{equation}
 {\rm Norm} = \frac{10^{-14}}{4 \pi [d_A(1+z)]^2 }\int n_e n_H dV
 \label{equ: APEC_norm}
\end{equation}
where $n_e$ and $n_H$ are the electron and ion number densities respectively (which are related by $n_e=1.17n_H$ in a fully ionized plasma, \citealt{Grevesse1998}), $d_A$ is the angular diameter distance to the source, and $z$ is the cluster redshift. From the deprojected APEC normalizations, we then calculated the density assuming spherical symmetry and a constant density in each shell, by calculating the projected volumes, V, of each shell in the 2D annuli.

\subsection{\textbf{\textit{Planck} analysis}}
\label{sec: Planck_analysis}
\begin{figure}
\begin{center}
\includegraphics[width=\columnwidth]{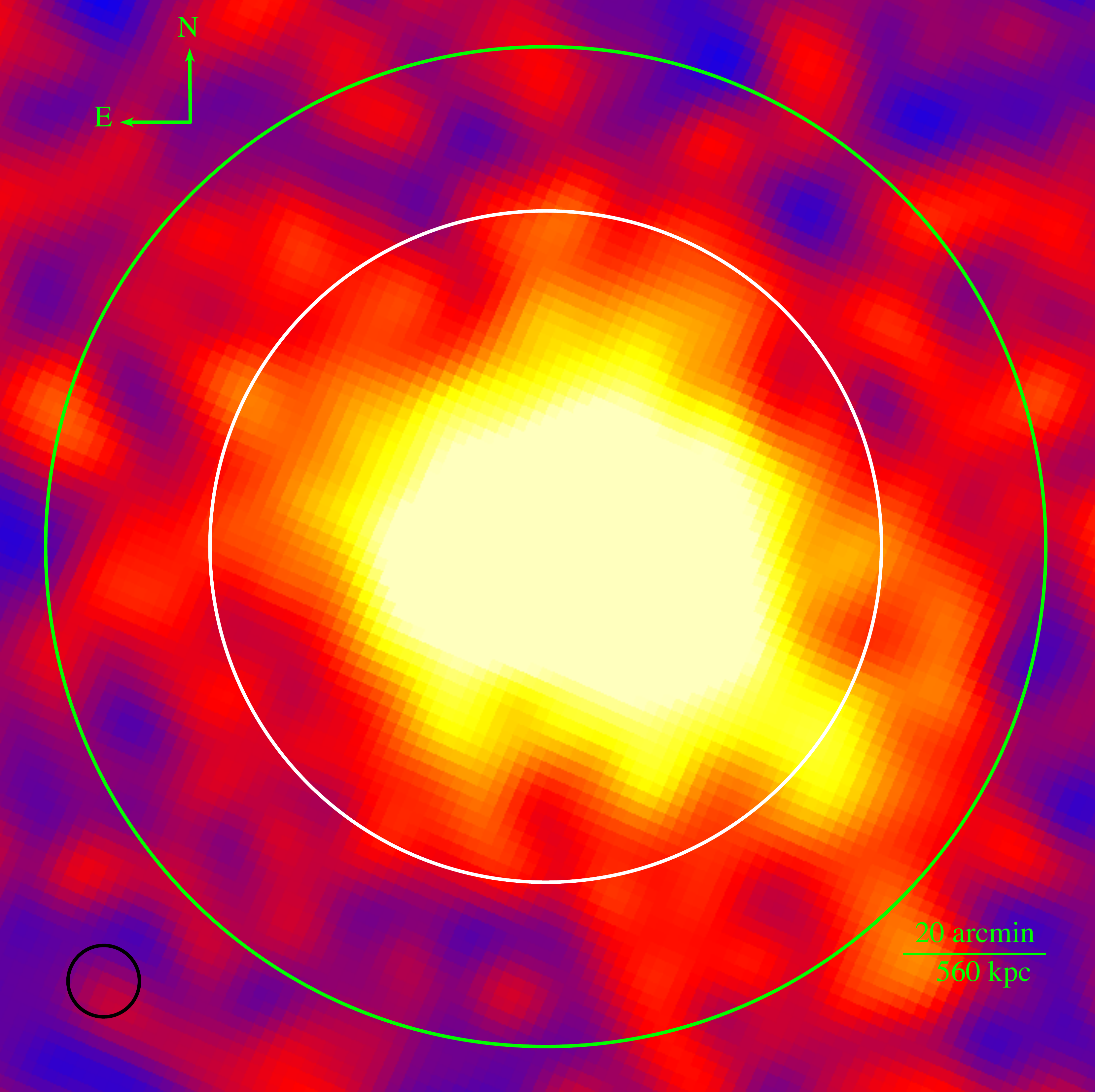}
\end{center}
\vspace{-0.2cm}
\caption{\textit{Planck} $y$ map of the Coma cluster, obtained by combining the \textit{Planck} frequency channels from 100 to 857 GHz. The white and green circles represent the location of $r_{500}$ and $r_{200}$, respectively. The black circle in the bottom left corner indicates the \textit{Planck} beam size of 10 arcmin, which corresponds to a physical size of 280 kpc.}
\label{fig: coma_sz}
\end{figure}

The \textit{Planck} data used in this work are prepared by \citet{ade2013planck}, using the Modified Internal Linear Combination Algorithm (MILCA). As in the analyses of the \citet{Tchernin2016}, \citet{ghirardini2018xmm}, and \citet{Ghirardini2019}, we use only the high frequency channels from 100 to 857 GHz owing to the much larger PSF of the low frequency Planck channels. Fig. \ref{fig: coma_sz} shows the Comptonization parameter $y$ map of the Coma cluster, obtained from a combination of \textit{Planck} channels from 100 to 857 GHz. The $y$ map has an effective PSF corresponding to 10 arcmin (FWHM), and a noise level of $2.3 \times 10^{-6}$. The \textit{Planck} $y$ map also shows the gas extends in the west and southwest directions, and the SZ effect signal is clearly detected towards the NGC 4839 substructure. Similar to the X-ray analysis, we excluded this region in our SZ analysis.

The dimensionless Comptonization parameter $y$, which determines the strength of interaction between the cosmic microwave background (CMB) photons and the hot ICM gas, is proportional to pressure, $P$, of the ICM gas integrated along the line of sight:
\begin{equation}
 y = \frac{\sigma_T}{m_e c^2}\int P(l)dl,
 \label{equ: y_parameter}
\end{equation}
where $\sigma_T$ is the Thomson scattering cross-section, $m_e$ is the electron mass, $c$ is the speed of light, and the integration is taken over the line of sight $l$.

Following \citet{ade2013planck}, the pressure profile of Coma was recovered by fitting the projected $y$ radial profile to the universal pressure formula proposed by \citet{Nagai2007}:
\begin{equation}
 P(x) = \frac{P_0}{(c_{500} x)^\gamma[1+(c_{500} x)^\alpha]^{(\beta-\gamma)/\alpha}},
 \label{equ: universal_pressure}
\end{equation}
where $x = r/r_{500}$, $P_0$ is the normalization parameter, $c_{500}$ is the concentration parameter defined at the characteristic radius $r_{500}$, and the parameters ($\gamma$, $\alpha$, $\beta$) are respectively slopes in the central ($x \ll 1/c_{500}$), intermediate ($x \sim 1/c_{500}$), and outer ($x \gg 1/c_{500}$) regions.

\begin{figure}
\begin{center}
\includegraphics[width=\columnwidth]{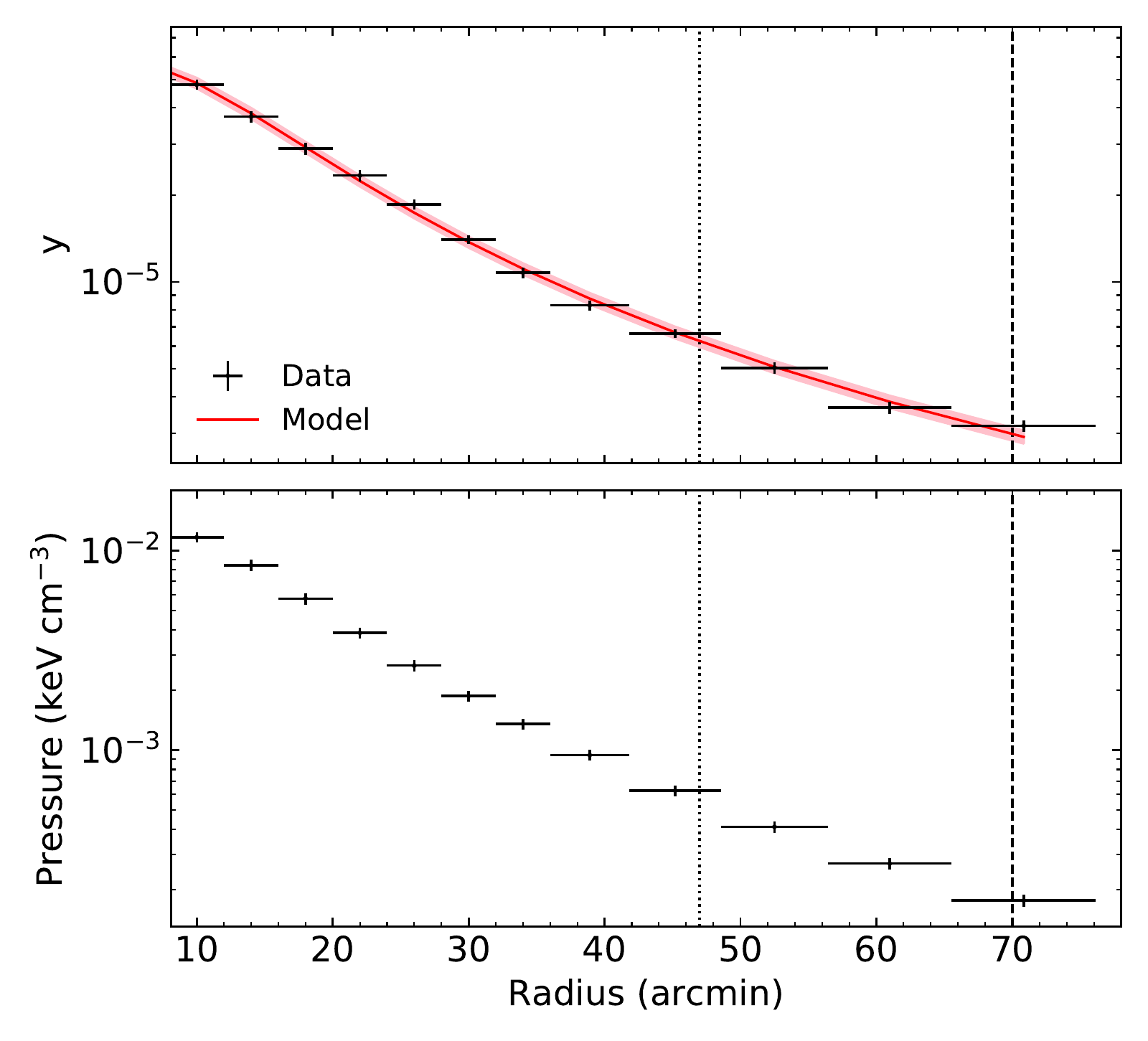}
\end{center}
\vspace{-0.5cm}
\caption{\textit{Top}: Azimuthally averaged $y$ profile of the Coma cluster with the best-fitting model. \textit{Bottom}: Pressure profile predicted from the best-fitting parameters. The shadow pink area indicates the 68.3 per cent confidence intervals obtained by MCMC simulations, and the vertical dotted and dashed lines represent, respectively, the $r_{500}$ and $r_{200}$ radii.
}
\label{fig: y_p_all}
\end{figure}

For the fitting processes, we used the affine-invariant ensemble sampler for Markov chain Monte Carlo implemented in the \textit{emcee} package \citep{goodman2010ensemble}. We let all five parameters in equation (\ref{equ: universal_pressure}) free to vary. Our estimated values of the best-fitting parameters are statistically consistent with those obtained from the best-fitting model, Model C, reported by \citet{ade2013planck}. The value of the reduced $\chi^2$ associated with this fit is 1.06 with 9 degrees of freedom. Fig. \ref{fig: y_p_all} presents the best-fitting $y$ model to the data and the pressure from the best-fitting parameters. We note that the model accurately reproduces the data over the entire radial range of the cluster. We take into account the covariance matrix between Planck data points, though this effect is small due to the large size of our bins.

\section{Joint X-ray and SZ effect analysis}
\label{sec: joint xray and sz}
The electron density profile recovered from the X-ray observations (Section \ref{sec: XMM_analysis}) and the pressure profile obtained from the SZ effect measurements (Section \ref{sec: Planck_analysis}) can be combined to obtain the gas temperature, entropy, masses, and gas mass fraction profiles.

\begin{figure}
\begin{center}
\includegraphics[width=\columnwidth]{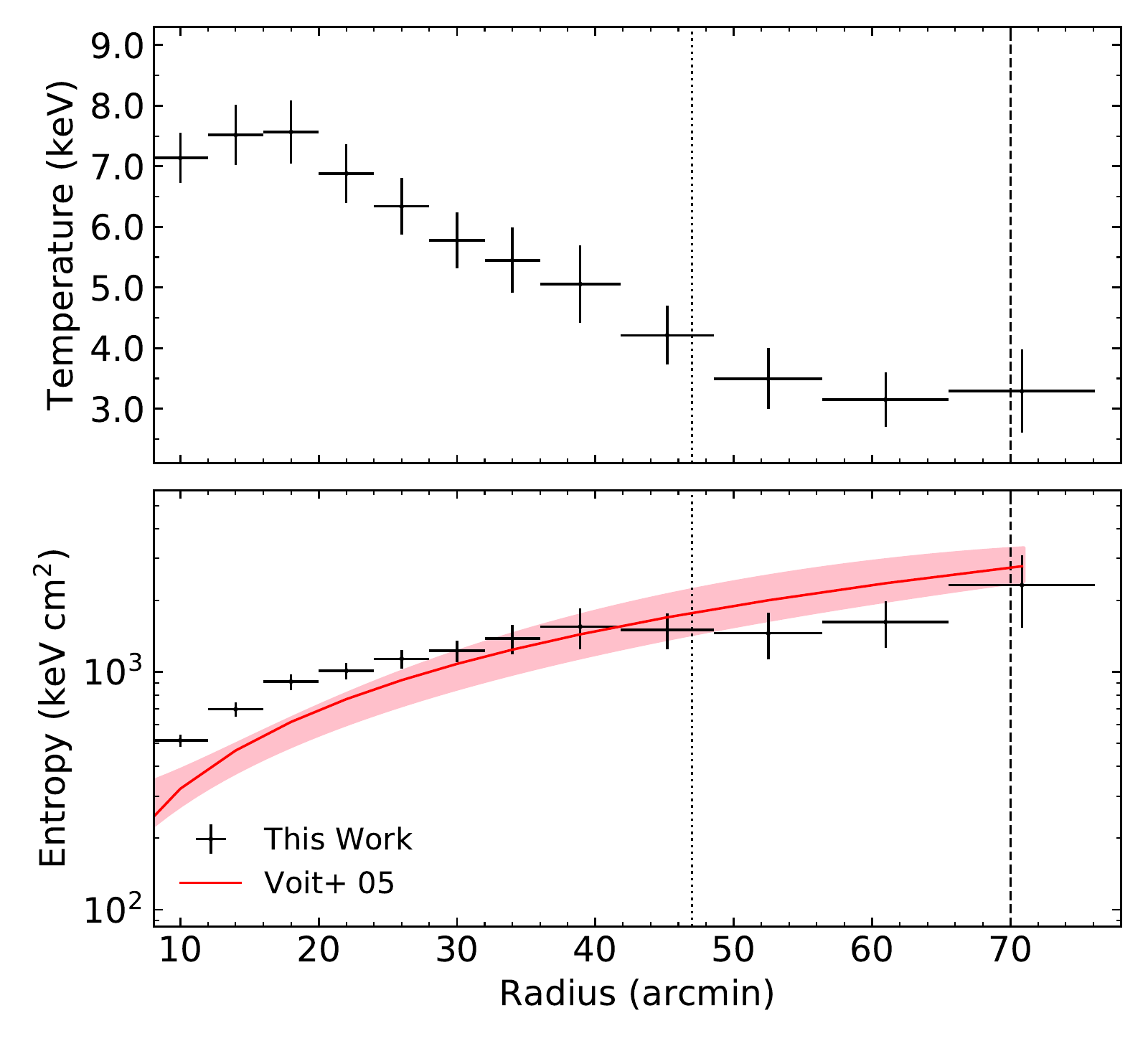}
\end{center}
\vspace{-0.5cm}
\caption{Temperature (\textit{top}) and entropy (\textit{bottom}) profiles of the Coma cluster obtained by combining the X-ray density and SZ pressure measurements. The red solid line indicates the entropy predicted by \citet{Voit2005}. The shaded pink area represents the region enclosed by the median profile and scatter of the non-radiative simulations in \citet{Voit2005}. The vertical dotted and dashed lines represent, respectively, the locations of the $r_{500}$ and $r_{200}$ radii. We plot using a linear radial axis to emphasize the outskirts.
}
\label{fig: temp_entropy_all}
\end{figure}

The temperature profile of the Coma cluster, as shown in the upper panel of Fig. \ref{fig: temp_entropy_all}, was recovered by dividing the pressure profile from \textit{Planck} by the electron density from the deprojection of the \textit{XMM-Newton} surface brightness profile. As we note, the temperature decreases from about 7.0 keV at a radius of 10 arcmin to about 3.0 keV at 70 arcmin.  

Similarly, we were able to recover the entropy profile through the gas pressure and density profiles as
\begin{equation}
 K=\frac{P}{n_e^{5/3}}.
 \label{equ: entropy}
\end{equation}
In the lower panel of Fig. \ref{fig: temp_entropy_all}, we present our measurements of the azimuthally averaged entropy profile of the Coma cluster.

The radial profile of the gas entropy is of particular interest since it tracks the thermal history of a galaxy cluster. Numerical simulations \citep{Voit2005} predicted that, in the absence of the radiative processes, the entropy profile outside the cluster core follows a power law with characteristic slope of 1.1,   
\begin{equation}
 \frac{K}{K_{500}}=1.47\bigg(\frac{r}{r_{500}}\bigg)^{1.1},
 \label{equ: Voit_entropy}
\end{equation}
where
\begin{equation}
 K_{500}=106\, {\rm{keV\, cm}}^2 \bigg(\frac{M_{500}}{10^{14}{\rm{M_\odot}}}\bigg)^{2/3}E(z)^{-2/3}f_b^{-2/3},
 \label{equ: K500}
\end{equation}
$M_{500}$ is the cluster mass at $r_{500}$, $E(z)$ describes the redshift evolution of Hubble parameter, and $f_b$ is the universal baryon fraction and assumed to be equal to 0.15 (following \citealt{Pratt2010}). 

The predicted entropy profile from this power-law relation is overplotted together with the measured entropy profile in the lower panel of Fig. \ref{fig: temp_entropy_all}. We see a clear excess in the entropy measured in the central region of the cluster out to a radius of about 30 arcmin. However, the recovered entropy is consistent with the expected trend at radii between 30-50 arcmin. Beyond 50 arcmin, although it has a shallower slope, the recovered entropy is statistically in agreement with the baseline entropy profile predicted by the numerical simulations \citep{Voit2005} given the scatter in the simulated entropy profiles.

To calculate the gas mass, we integrated the density profile over a given volume, and then multiplied by the mean mass per electron, $\mu_{\rm{e}} m_{\rm{p}}$. The gas mass can be expressed as  
\begin{equation}
 M_{\rm{gas}}=4\pi \mu_{\rm{e}} m_{\rm{p}} \int n_{\rm{e}}(r)r^2dr,
 \label{equ: gas_mass}
\end{equation}
where the limit is taken out to some cut-off radius. The total mass of the cluster, which is mainly made up of the dark matter, can be deduced from the gas structure. Its distribution relates to the pressure profile of the ICM gas. Assuming that the gas is in the hydrostatic equilibrium in the gravitational potential well of the cluster with no significant flow of matter, the total mass takes the form:
\begin{equation}
 M_{\rm{tot}}=-\frac{r^2}{G \mu_{\rm{e}} m_{\rm{p}} n_{\rm{e}}(r)} \frac{{\rm{d}}P(r)}{{\rm{d}}r},
 \label{equ: tot_mass}
\end{equation}
where $G$ is the Newtonian gravitational constant, and the pressure term satisfies the ideal gas law. 

Both gas and total masses were calculated using the following method. We used a four-point spline interpolation to compute the integral in equation (\ref{equ: gas_mass}), and a three-point quadratic interpolation for the pressure derivative in equation (\ref{equ: tot_mass}). The gas mass fraction is then recovered as $f_{\rm{gas}}=M_{\rm{gas}}/M_{\rm{tot}}$. 

\begin{figure}
\begin{center}
\includegraphics[width=\columnwidth]{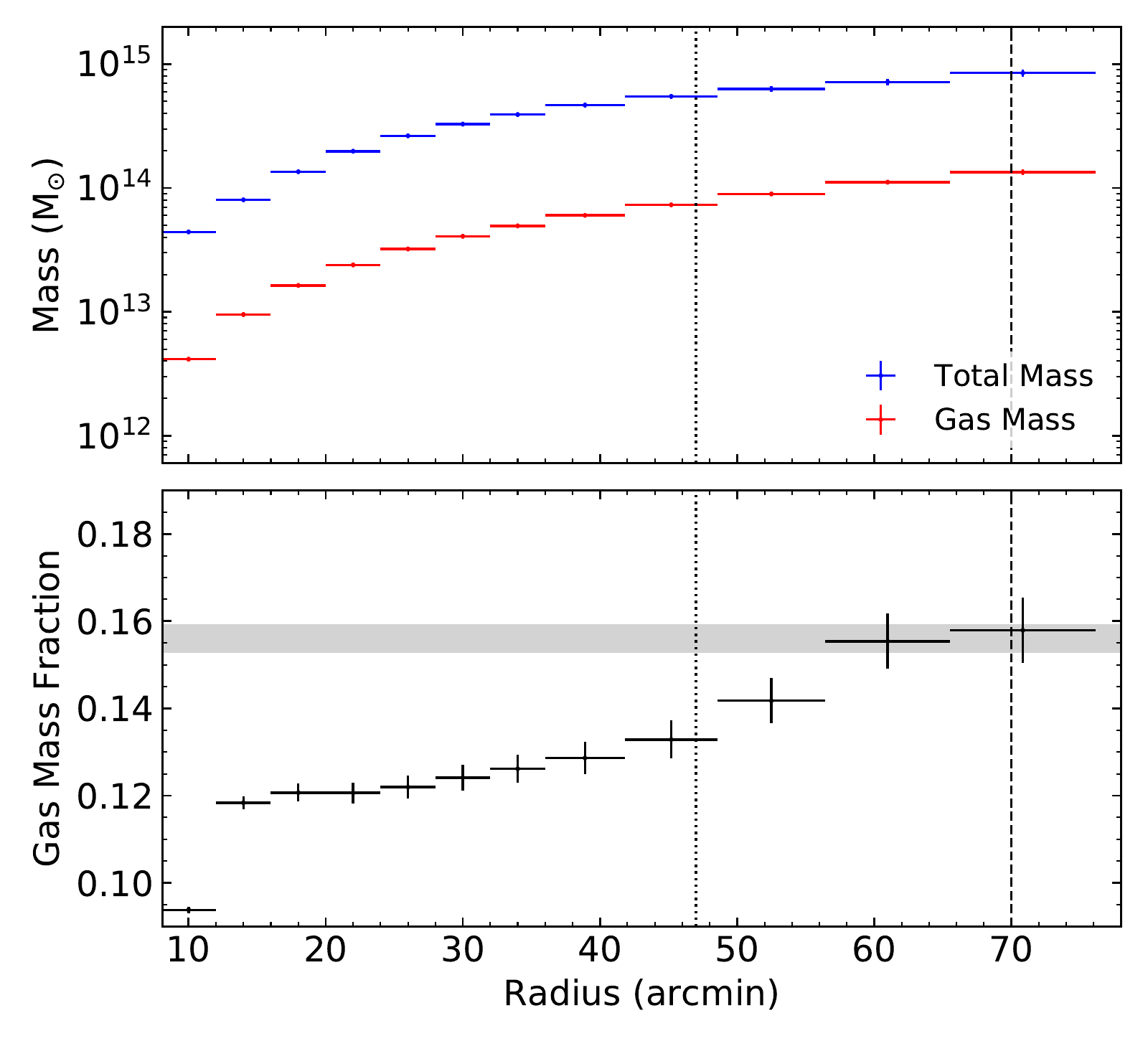}
\end{center}
\vspace{-0.5cm}
\caption{\textit{Top}: Gas and total mass profiles of the Coma cluster. \textit{Bottom}: Azimuthally averaged gas mass fraction of Coma. The gray shadow area represents the \textit{Planck} universal baryon fraction \citep{ade2016planck}. The vertical dotted and dashed lines mark the positions of the $r_{500}$ and $r_{200}$ radii, respectively.
}
\label{fig: masses_&_fraction_all}
\end{figure}

Fig. \ref{fig: masses_&_fraction_all} presents the gas mass, total mass, and gas mass fraction profiles of the Coma cluster out to the virial radius. As a general trend, these measurements increase radially with increasing radius, and reach the maximum value at the cluster outskirts. The estimated total mass of Coma at $r_{200}$ ($=2.0$ Mpc) is $(8.50 \pm 0.55) \times 10^{14} \rm{M}_\odot$. This value in excellent agreement with the $M_{200}$ value from the $Y_X-M$ scaling relation obtained in \citet{ade2013planck} of $8.6 \times 10^{14} \rm{M}_\odot$. Our measurements for the azimuthally averaged gas mass fraction at $r_{200}$ are also statistically in agreement with the universal baryon fraction of 0.156 $\pm$ 0.003 reported by \citet{ade2016planck}. 

\begin{figure}
\begin{center}
\includegraphics[width=\columnwidth]{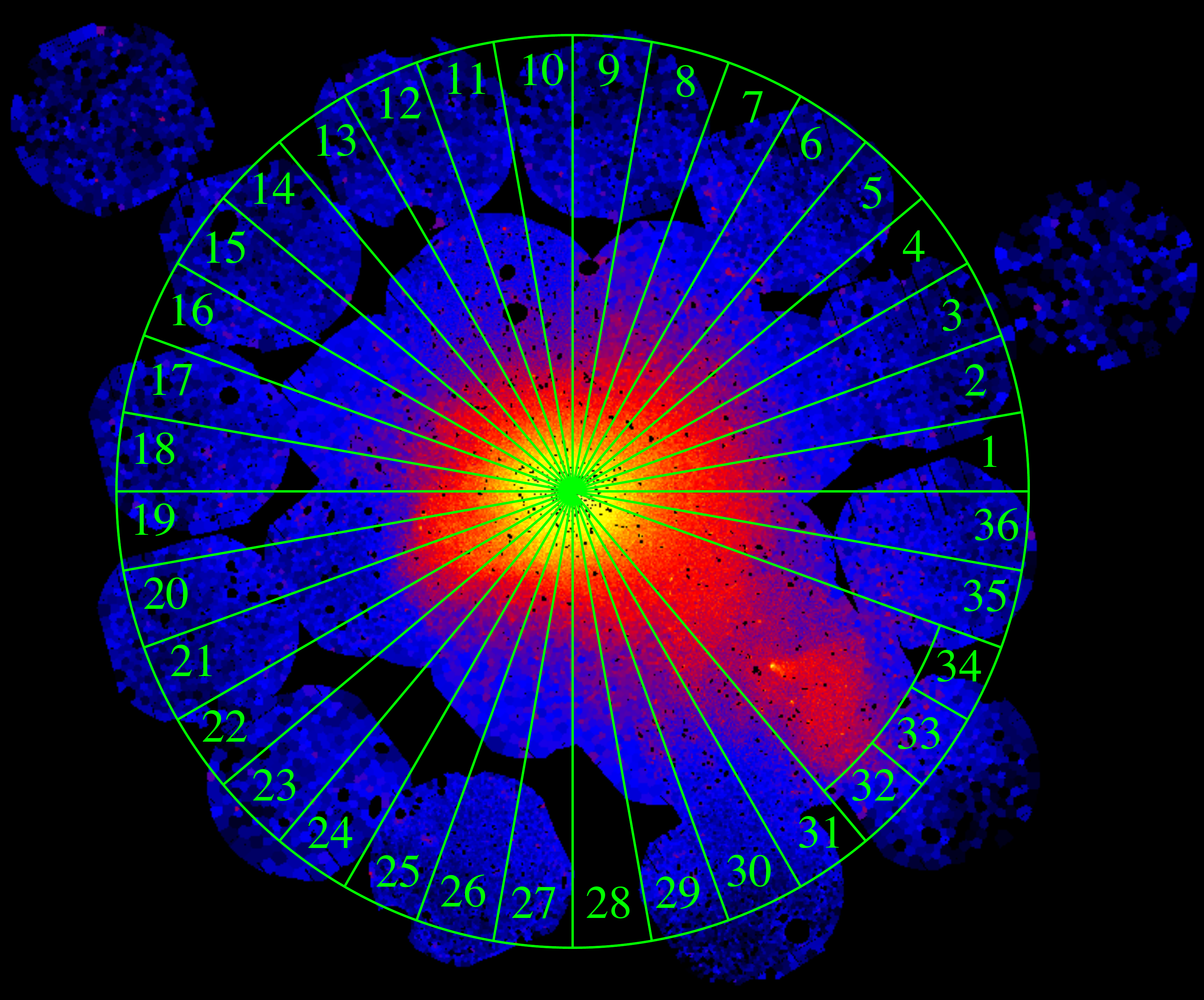}

\end{center}
\vspace{-0.3cm}
\caption{Locations of the 36 azimuthal sectors overlaid on the X-ray image of the Coma cluster. Similar to the azimuthally averaged analysis, we excluded the region in the southwest direction where the substructure NGC 4839 is located. Any gaps in the mosaic have been linearly interpolated over, and the values for the missing sector 28 are tied to those of sector 29.
  }
\label{fig: coma_sectors}
\end{figure}
\section{Azimuthal Sector Analysis}
\label{sec: sector analysis}
Taking advantage of our high-quality data from both X-ray and SZ effect observations, we can apply the analysis described in the previous sections on smaller regions of the Coma cluster. \citet{ghirardini2018xmm} found when performing their joint XMM and Planck analysis on the cluster Abell 2319 (which has $r_{200}=32$ arcmin) that they could divide Abell 2319 into 8 sectors to assess the azimuthal variations in thermodynamic properties. Coma's much larger spatial extent allows us to dramatically increase the number of sectors into which it can be resolved by \textit{Planck}. The Coma cluster (with $r_{200}=70$ arcmin) covers an area that is $(70/32)^2 = 4.8$ times greater than that of Abell 2319, so we can increase the number of sectors in our analysis of the azimuthal variations in Coma by this factor to $8\times4.8=38.4$. For simplicity, and to make the presentation of our results intuitive, we divide our analysis in 36 sectors (since there are 360 degrees in a circle). We find 36 to be the optimum number of sectors, as having more sectors would mean we have too few Voronoi tesselated regions to measure the clumping factor accurately in each sector, while using fewer sectors would be under-utilizing the data. When binning each sector radially to form profiles, we ensure that we have enough independent Voronoi tessellated regions in each radial bin we study to accurately measure the clumping bias (as discussed earlier in Section \ref{sec: XMM_analysis}). In the small number of sectors where the XMM coverage in the outskirts is low (sectors 1 and 4 and 34), we use the clumping values from their neighbouring sectors. The analysis in azimuthal sectors allows us to measure the variations in the thermodynamic properties of the Coma cluster out to the virial region. In this analysis of the sectors, any small gaps in the XMM mosaic are linearly interpolated over.

Fig. \ref{fig: coma_sectors} shows the locations of the azimuthal sectors where the X-ray and SZ effect data were extracted and analysed. Each of the 36 azimuthal sectors has an angular width of 10 degrees and extends up to the $r_{200}$ radius. For sector 28, where there is no X-ray coverage in the outskirts, we tie the X-ray surface brightness to be that same as that in the adjacent sector 29. Along the southwest direction, where the bright subgroup NGC 4839 is located, we excluded the region that covers the cluster core up to a radius of 60 arcmin, so as to exclude the emission from NGC 4839 and its tail. The recovered temperature and entropy along this direction outside 60 arcmin (sectors 32, 33, and 34) are shown in Fig. \ref{fig: entropy_temp_last_bin}. All measurements presented in this section take account of the systematic uncertainties discussed in Section \ref{sec: XMM_analysis}, and corrected for gas clumping (see Fig. \ref{fig: clumping}).  

The density profiles in the azimuthal sectors, as shown in Fig. \ref{fig: ne_sectors}, exhibit a small scatter from the averaged density profile in the central region of the cluster, but scatter is more significant outward towards the cluster outskirts, agreeing with the prediction of cosmological simulations \citep[e.g.][]{nelson2014weighing}. Some azimuthal sectors show an excess gas density in the outskirts region beyond $r_{500}$, compared to the azimuthally averaged density profile. This excess, however, is significantly higher beyond a radius of 40 arcmin in sectors that lie between south and west. Apart from this excess, the gas density profiles in some sectors that lie between east and south drop sharply between 30-40 arcmin and then flat out at radii between 40-50 arcmin. As found in \citet{brown2011diffuse}, the edge of this sharp drop in the gas density agrees well with the outermost edge of the giant radio halo observed with the Westerbork Synthesis Radio Telescope (WSRT, shown in Fig. \ref{fig: coma_xray_radio}).

The pressure profiles in the azimuthal sectors, obtained by fitting the projected $y$ profile in each sector to the universal pressure profile (equation \ref{equ: universal_pressure}), are shown in Fig. \ref{fig: pressure_sectors}. Like the density profile, we see a small scatter from the averaged pressure profile in the central region of the cluster, but increases radially with the cluster radius, although smaller than that for the density profiles. 

\begin{figure*}
\begin{center}

\includegraphics[width=0.9\textwidth]{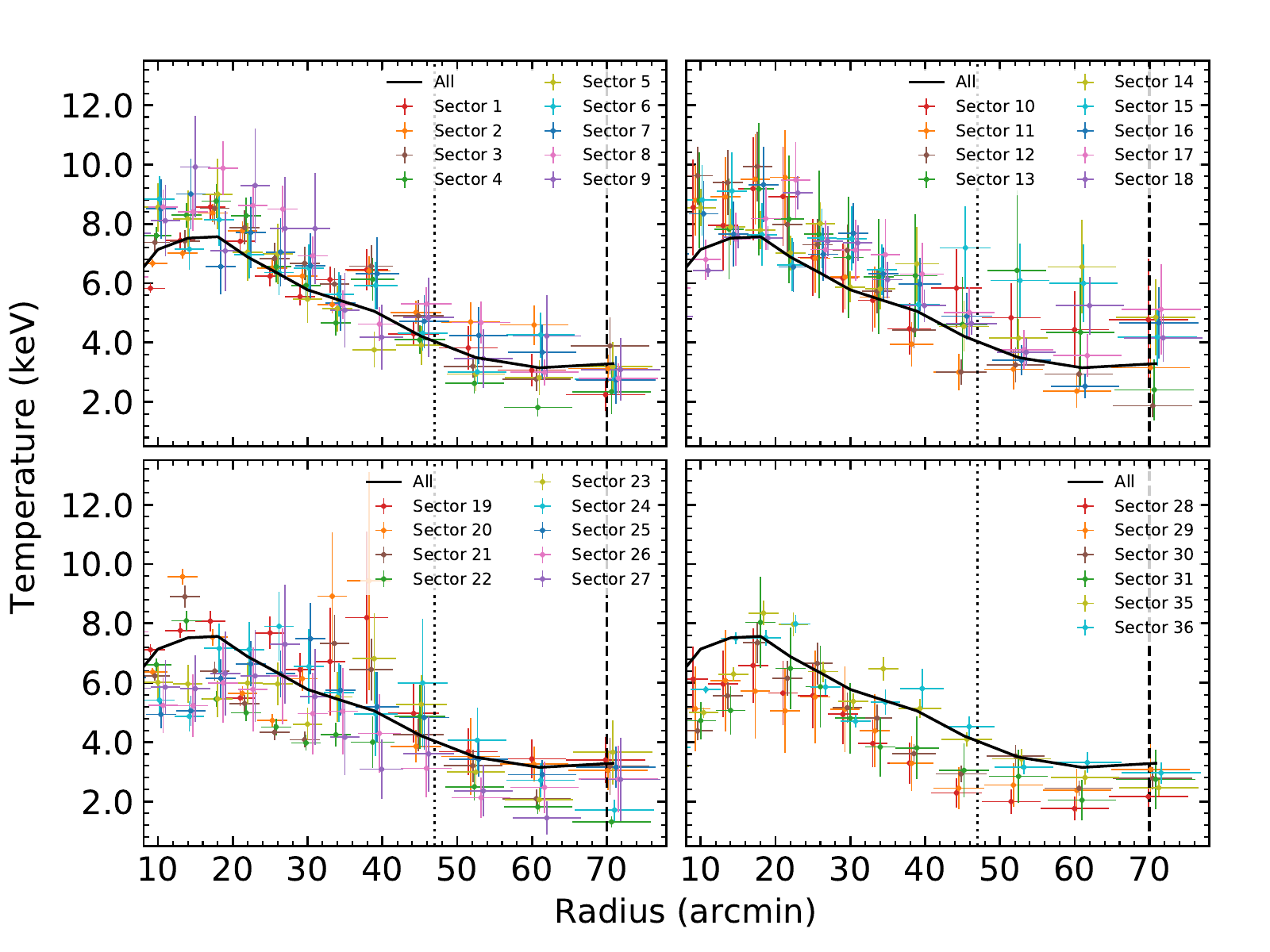}

\end{center}
\vspace{-0.5cm}
\caption{Temperature profiles in the azimuthal sectors of the Coma cluster obtained from combining the X-ray density and SZ pressure measurements. The error bars are the 1 $\sigma$ percentiles computed using a Monte Carlo technique. The thick black line shows the azimuthally averaged temperature profile. The vertical dotted and dashed lines mark the positions of the $r_{500}$ and $r_{200}$ radii, respectively. We introduce small offsets between data points at the same radius to aid readability.
  }
\label{fig: temp_sectors}
\vspace{-0.3cm}
\end{figure*}
\begin{figure}
\begin{center}
\hbox{
\includegraphics[width=0.45\textwidth]{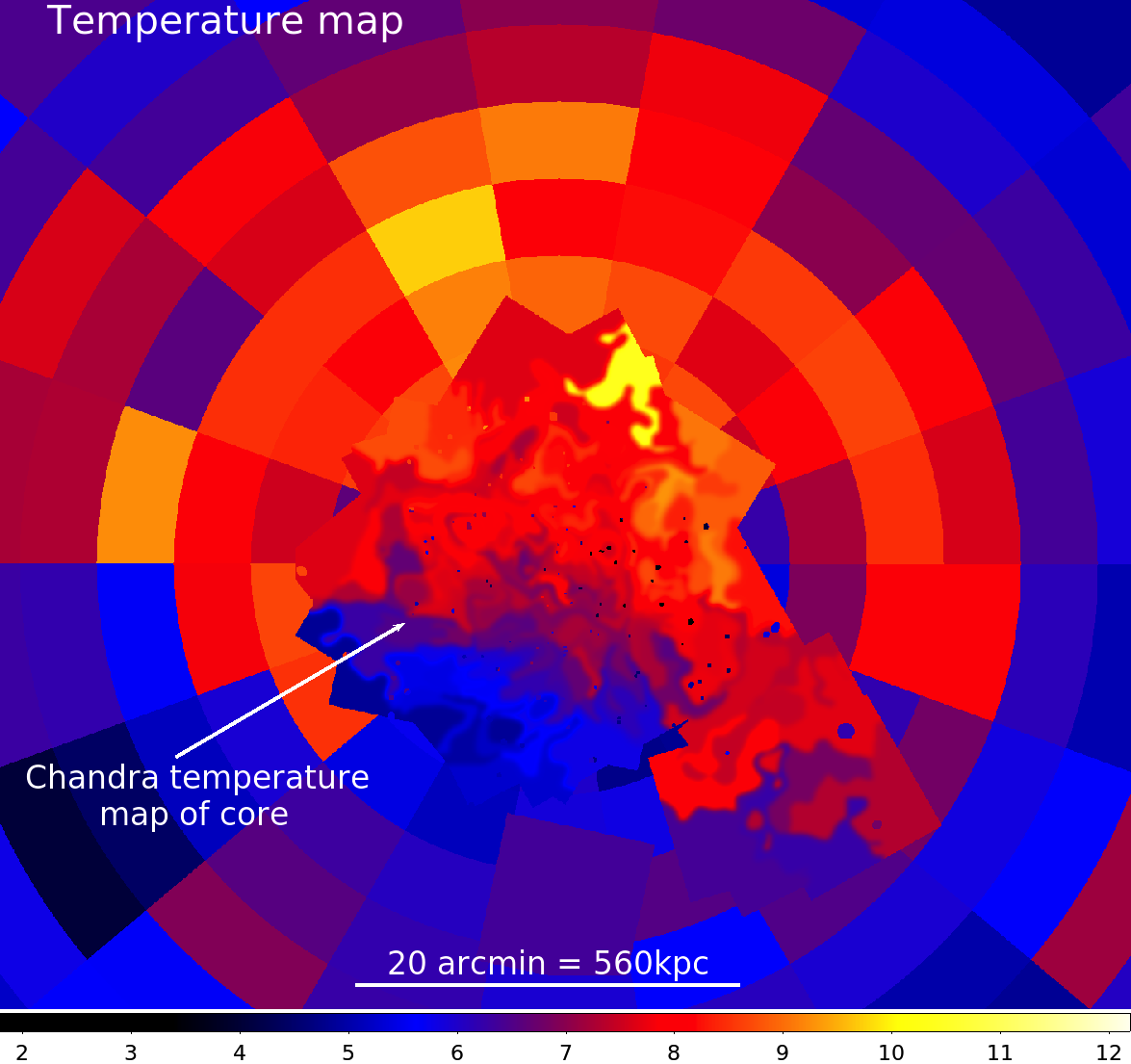}

}
\end{center}
\vspace{-0.5cm}
\caption{Temperature map of the central regions of Coma from the joint \textit{XMM/Planck} analysis (sectors) compared to the \textit{Chandra} temperature map of the core from \citet{sanders2013linear}. The asymmetry between the hotter northern region and cooler southern region found in the core with \textit{Chandra} is found to extend out to much larger radius. 
 }
\label{fig: temp_entropy_map}
\end{figure}

In Fig. \ref{fig: temp_sectors}, we show the temperature profiles in the azimuthal sectors recovered from the density and pressure measurements. In this figure, we also show the azimuthally averaged temperature profile. As a general trend, the temperature profiles in most azimuthal sectors exhibit broadly consistent radial trends, with the gas temperature dropping from about 8.0 keV at a radius of 10 arcmin to about 3.0 keV in the cluster outskirts, with few exceptions. For instance, the temperature measurements in sectors 10-18 (the north east of the cluster)  are, on average, higher than the azimuthally averaged profile over the most cluster radial range. In contrast, in the azimuthal sectors that lie between east and south, the temperature measurements are, on average, lower than the azimuthally averaged profile from the cluster centre out to the cluster outskirts. A temperature map of the central 40 arcmin of Coma region is shown in Fig. \ref{fig: temp_entropy_map}, showing the north/south asymmetry in the temperature. Our results agree well with the hot north and colder south observed in the central regions with \textit{Chandra} in \citet{sanders2013linear}, and show that this asymmetry extends to much around 30-40 arcmin. 

In sectors to the south-west, the temperature profiles are statistically below the azimuthally averaged temperature profile from the centre of the cluster out to about 20 arcmin. From about 20-50 arcmin, the temperature profiles show a drop from about 6.0 keV to about 2.0 keV. Beyond $r_{500}$, however, the temperature profiles along these sectors decline only slightly with radius or remain unchanged.

\begin{figure*}
\begin{center}

\includegraphics[width=0.85\textwidth]{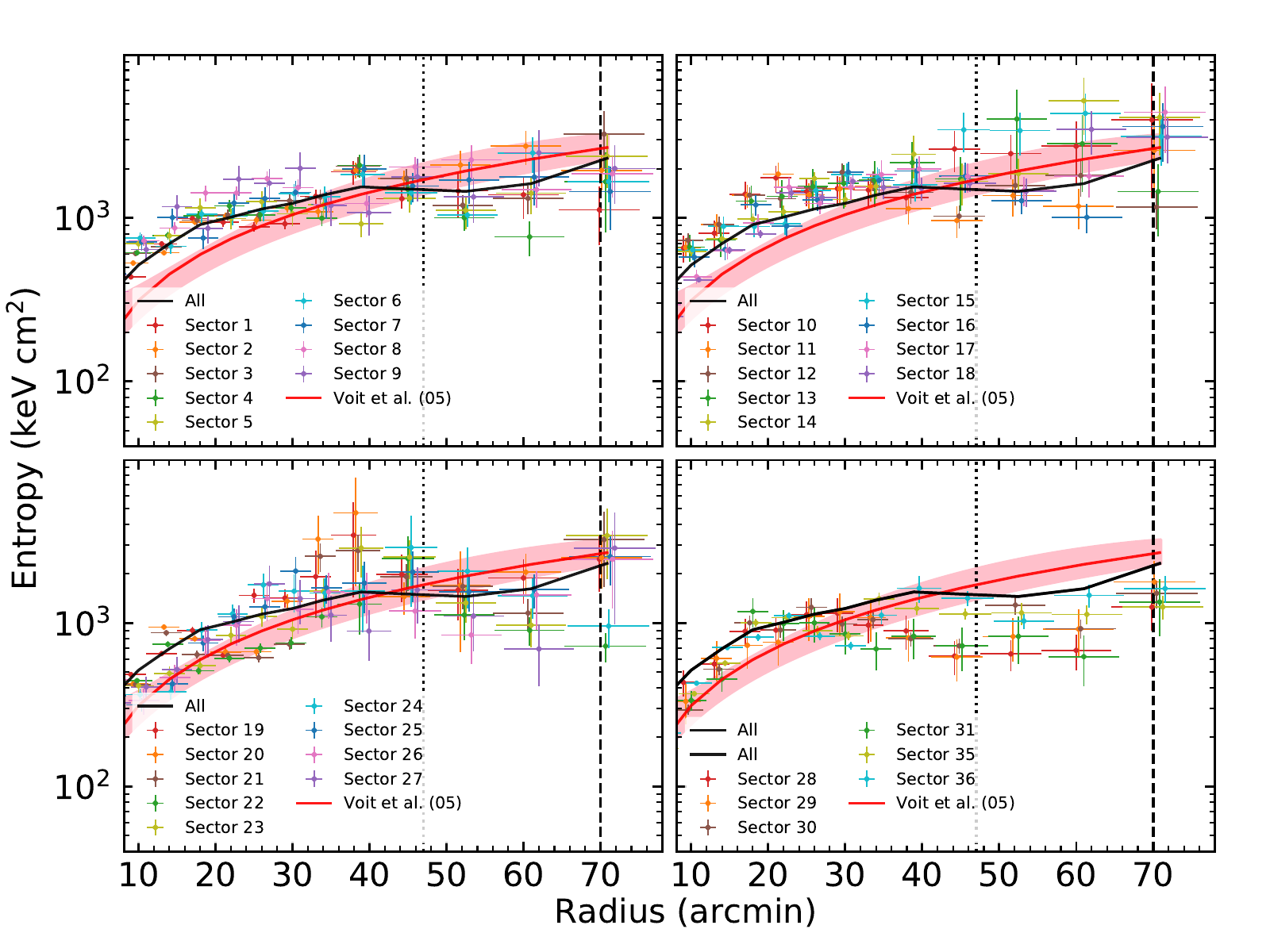}
\end{center}
\vspace{-0.5cm}
\caption{Same as Fig. \ref{fig: temp_sectors}, except for entropy. The red solid line indicates the entropy predicted by \citet{Voit2005}. The shaded pink area represents the region enclosed by the median profile and scatter of the non-radiative simulations in \citet{Voit2005}.
  }
\label{fig: entropy_sectors}
\vspace{-0.3cm}
\end{figure*}

 The entropy profiles in the azimuthal sectors are shown in Fig. \ref{fig: entropy_sectors}. We plot these profiles using a linear radial axis to emphasise the outskirts regions. In this figure, we show the azimuthally averaged profile along with the power-law entropy profile predicted by non-radiative simulations \citep{Voit2005}. Within a radius of 30 arcmin, the entropy profiles along most azimuthal sectors, on average, exhibit an apparent excess with respect to the power-law profile predicted by non-radiative simulations. Beyond a radius of 30 arcmin, the entropy profiles show a better agreement with the power-law behavior predicted from simulations. However, in the south-western quarter of the cluster, the entropy flattens and deviates significantly below the predicted profile. For sectors that lie between north and east, in contrast, we see an apparent entropy excess along several sectors out to the outskirts, implying the presence of merger activities which is likely associated with infalling subgroups \citep{simionescu2013thermodynamics}. 
 
\begin{figure}
\begin{center}
\hbox{

\includegraphics[width=0.45\textwidth]{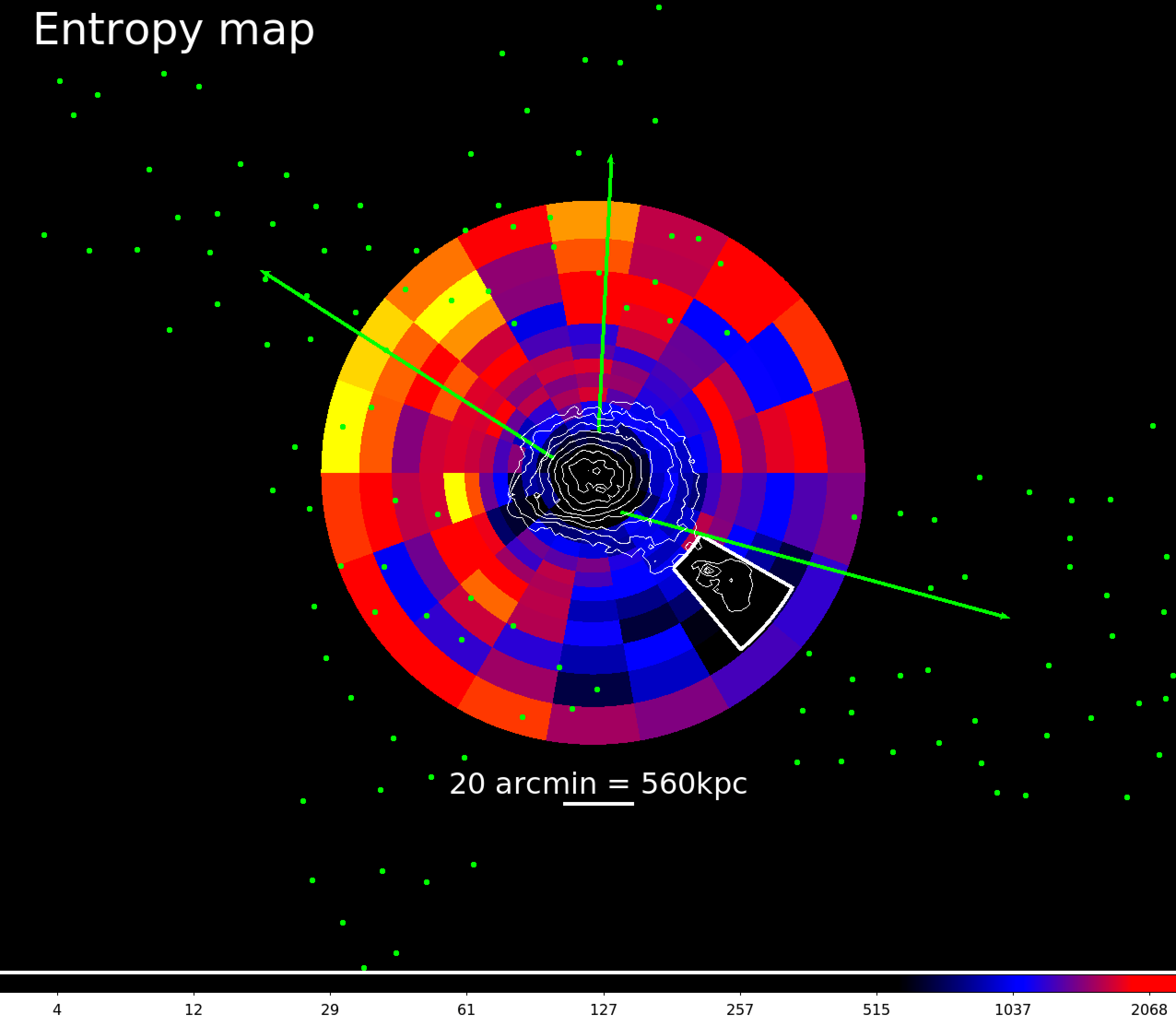}
}
\end{center}
\vspace{-0.5cm}
\caption{Wide-scale entropy map of the Coma cluster from the joint \textit{XMM/Planck} analysis. The white contours show the brightest X-ray features. The thick white sector around the merging group NGC 4839 is excluded from the map. The green arrows show the direction of the cosmic web filaments found in \citet{malavasi2020like} that connect the Coma cluster to neighbouring galaxy clusters. The filament galaxies from \citet{Mahajan2018} are shown as the green dots. We see that the south western area has systematically lower entropy than the rest of the cluster, possibly indicating the presence of a filamentary gas stream of low entropy gas, extending deep into the Coma cluster in this direction. Any gaps in the mosaic have been linearly interpolated over, and the values for the missing sector 28 are tied to those of sector 29.
 }
\label{fig: temp_entropy_map2}
\vspace{-0.3cm}
\end{figure}
 
 The entropy map of the outskirts is shown in Fig. \ref{fig: temp_entropy_map2}. The low entropy region to the south west coincides with the western filament connecting the Coma cluster to the cosmic web found in \citet{malavasi2020like}, indicated by the green arrow in this figure. The locations of the filament galaxies found in \cite{Mahajan2018} are plotted as the green points on Fig. \ref{fig: temp_entropy_map2}, showing a strong correlation between the size and location of this cosmic web filament, which has a width around 2 Mpc, and the entropy decrement in the south western quadrant.

\begin{figure*}
\begin{center}

\includegraphics[width=0.8\textwidth]{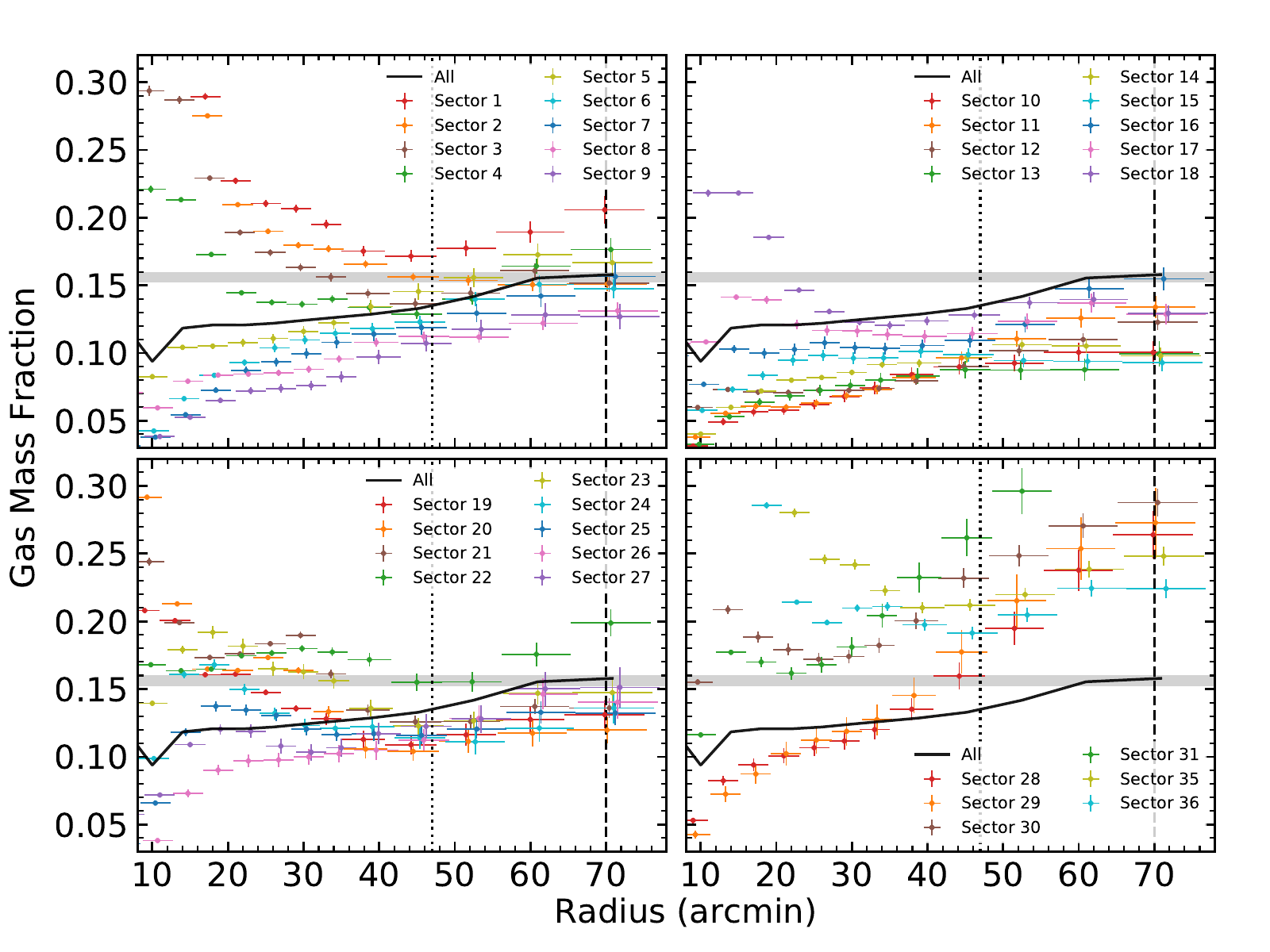}

\end{center}
\vspace{-0.5cm}
\caption{Same as Fig. \ref{fig: temp_sectors}, except for the gas mass fraction. The gray shadow area represents the \textit{Planck} universal baryon fraction \citep{ade2016planck}. 
  }
\label{fig: gas_mass_fraction_sectors}
\vspace{-0.3cm}
\end{figure*}

We also computed the gas and total mass profiles in the azimuthal sectors using the same methods (see Figs. \ref{fig: gas_mass_sectors} and \ref{fig: tot_mass_sectors}). From these mass measurements, the gas mass fraction profile is then recovered for all azimuthal sectors. In Fig. \ref{fig: gas_mass_fraction_sectors}, we show the gas mass fraction profiles in the azimuthal sectors, along with the azimuthally averaged profile and the universal baryon fraction measurement reported by \citet{ade2016planck}. As a general trend, the results indicate that the computed gas mass fractions increase slightly with a radius within the central and intermediate radii of the cluster and then tend to flat out at larger radii. However, this is not the case in every azimuthal sector. For example, the gas mass fraction measurements in some sectors that lie between north and west are significantly high at the cluster centre, indicating that the total mass in these sectors may be underestimated as a result of the presence of the possible non-thermal pressure component in the cluster core. However, the gas mass fraction profiles along these sectors drop outward, in better agreement with the universal baryon fraction at larger radii. In the south western quadrant, the gas mass fraction in the outskirts is found to be far higher, well in excess of the mean cosmic baryon fraction.   

\subsection{Azimuthal scatter profiles}
\begin{figure}
\begin{center}
\includegraphics[width=1.0\linewidth]{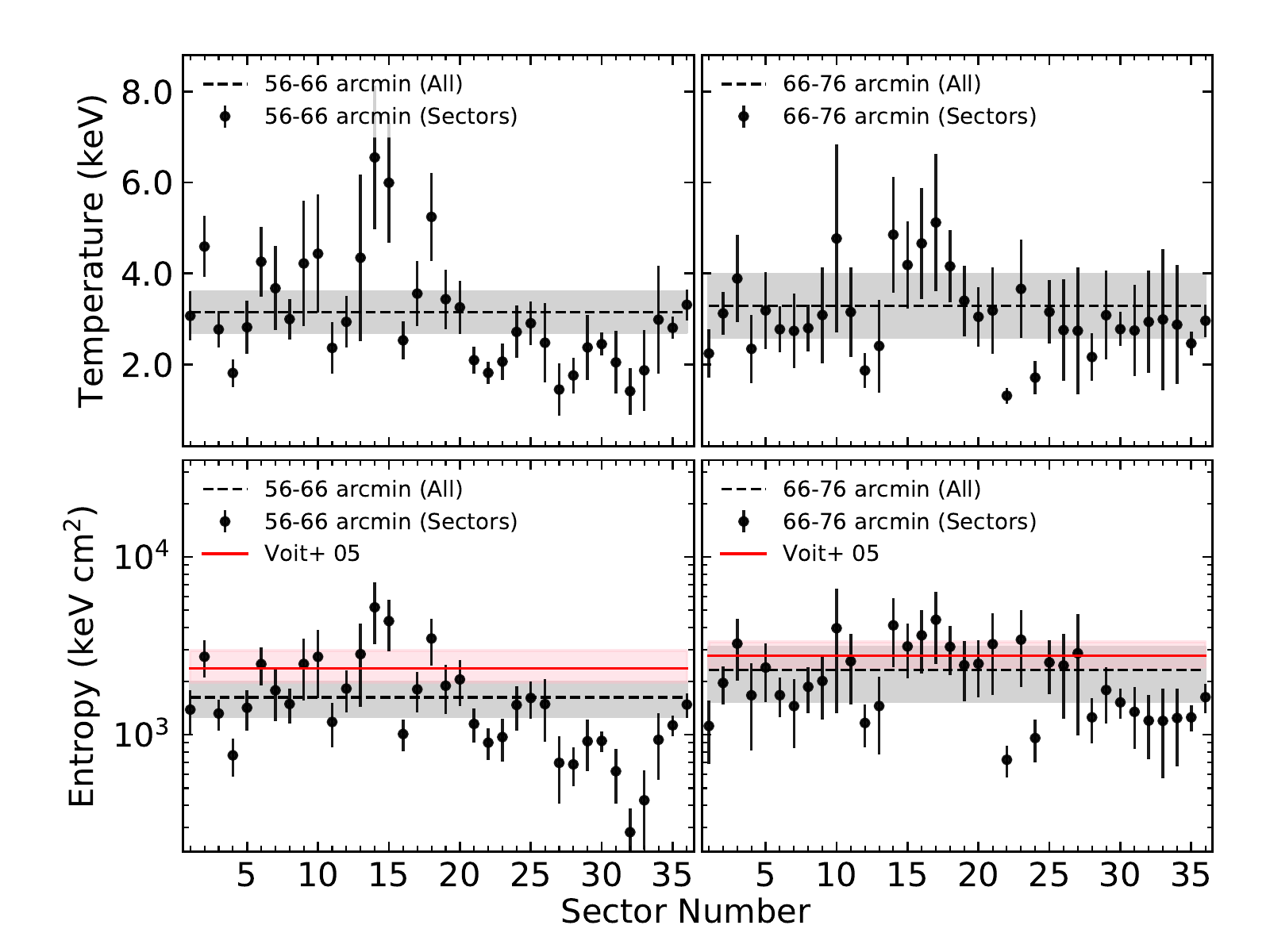}
\end{center}
\vspace{-0.5cm}
\caption{Azimuthal variation in the gas temperature (\textit{top}) and entropy (\textit{bottom}) in the last two annuli for all 36 azimuthal sectors. The dotted lines are the azimuthally averaged temperature (\textit{top}) and entropy (\textit{bottom}), omitting the sectors concident with the merging group NGC 4839 (sectors 32 to 34). The shaded gray regions represent 1$\sigma$ uncertainty. The red solid line indicates the entropy predicted by \citet{Voit2005}. The shaded pink region corresponds to the region enclosed by the median profile and scatter of the non-radiative simulations in \citet{Voit2005}.}
\label{fig: entropy_temp_last_bin}
\vspace{-0.3cm}
\end{figure}

In Fig. \ref{fig: entropy_temp_last_bin}, we plot the azimuthal variation of the temperature and entropy in the outermost two annuli. We see clear azimuthal variation in the entropy, which is systematically lower in the south west direction, between sectors 27 and 36.

To investigate the level of asymmetry in the radial profiles of the thermodynamic quantities recovered from X-ray and SZ effect observations in more detail, we computed the azimuthal scatter \citep{vazza2011scatter,ghirardini2018xmm} in all sectors with respect to the azimuthally averaged profile. For a given number of sectors $N$, the azimuthal scatter is defined at each radius $r$ as
\begin{equation}
 \sigma_q(r)=\sqrt{ \frac{1}{N}\sum\limits_{i=1}^N\bigg(\frac{q_i(r)-\bar{q}(r)}{\bar{q}(r)}\bigg)^2},
 \label{equ: azimuthal_scatter}
\end{equation}
where $q_i(r)=\{C, n_e, P, T, K, M_{\rm{gas}}, M_{\rm{tot}}, f_{\rm{gas}}\}$ in sector $i$, and $\bar{q}(r)$ is the azimuthally averaged profile taken over the entire cluster volume.

\begin{figure}
\begin{center}

\includegraphics[width=\columnwidth]{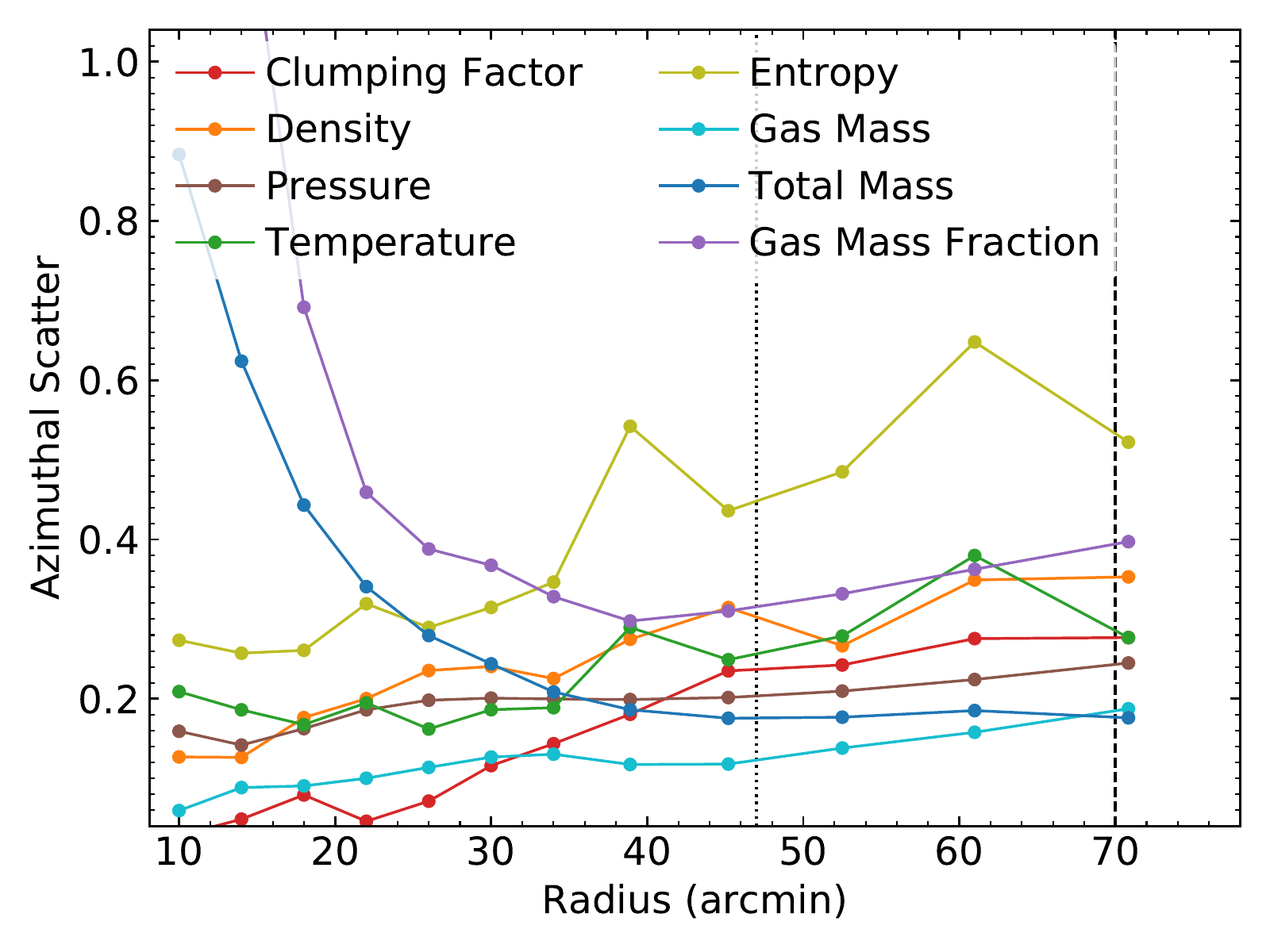}

\end{center}
\vspace{-0.5cm}
\caption{Azimuthal scatter in the radial profiles of the thermodynamic quantities of the Coma cluster recovered from X-ray and SZ effect observations. A small value of azimuthal scatter means no significant deviations from spherical symmetry, while a large azimuthal scatter suggests an asymmetric gas. The vertical dotted and dashed lines show the positions of the $r_{500}$ and $r_{200}$ radii, respectively.
  }
\label{fig: scatter_plot}
\vspace{-0.3cm}
\end{figure}

The scatter in the radial profiles of the recovered thermodynamic quantities of the Coma cluster is shown in Fig. \ref{fig: scatter_plot}. A small value of $\sigma_q(r)$ implies no significant deviations from spherical symmetry, while a large $\sigma_q(r)$ indicates a very asymmetric gas. Overall, the azimuthal scatter $\sigma_q(r)$ of the recovered thermodynamic quantities increases towards the cluster outskirts, in agreement with the predictions of cosmological simulations. The scatter in the temperature and density in the $r_{500}$ to $r_{200}$ is in the range 0.2-0.3, in good agreement with the range found in the simulations of \citet{vazza2011scatter}.

Although this is a general trend, there are a few exceptions. For example, the gas mass fraction exhibits a significant scatter in the central region of the cluster. Such large scatter might indicate that the total mass is underestimated within the central region of the cluster, supporting the presence of significant non-thermal pressure component in the core. The presence of such pressure is highly likely in this non-cool-core system and has previously been reported \citep[see][for example]{neumann2003dynamical,schuecker2004probing,simionescu2013thermodynamics}. Furthermore, there is a particular location near a radius of 40 arcmin where we observe a significant jump in the azimuthal scatter of the gas temperature and entropy. This location coincides well with the outermost edge of the giant radio halo observed at 352 MHz with the WSRT.

\subsection{Comparison with \textit{Suzaku}}

\begin{figure}
\begin{center}

\includegraphics[width=\columnwidth]{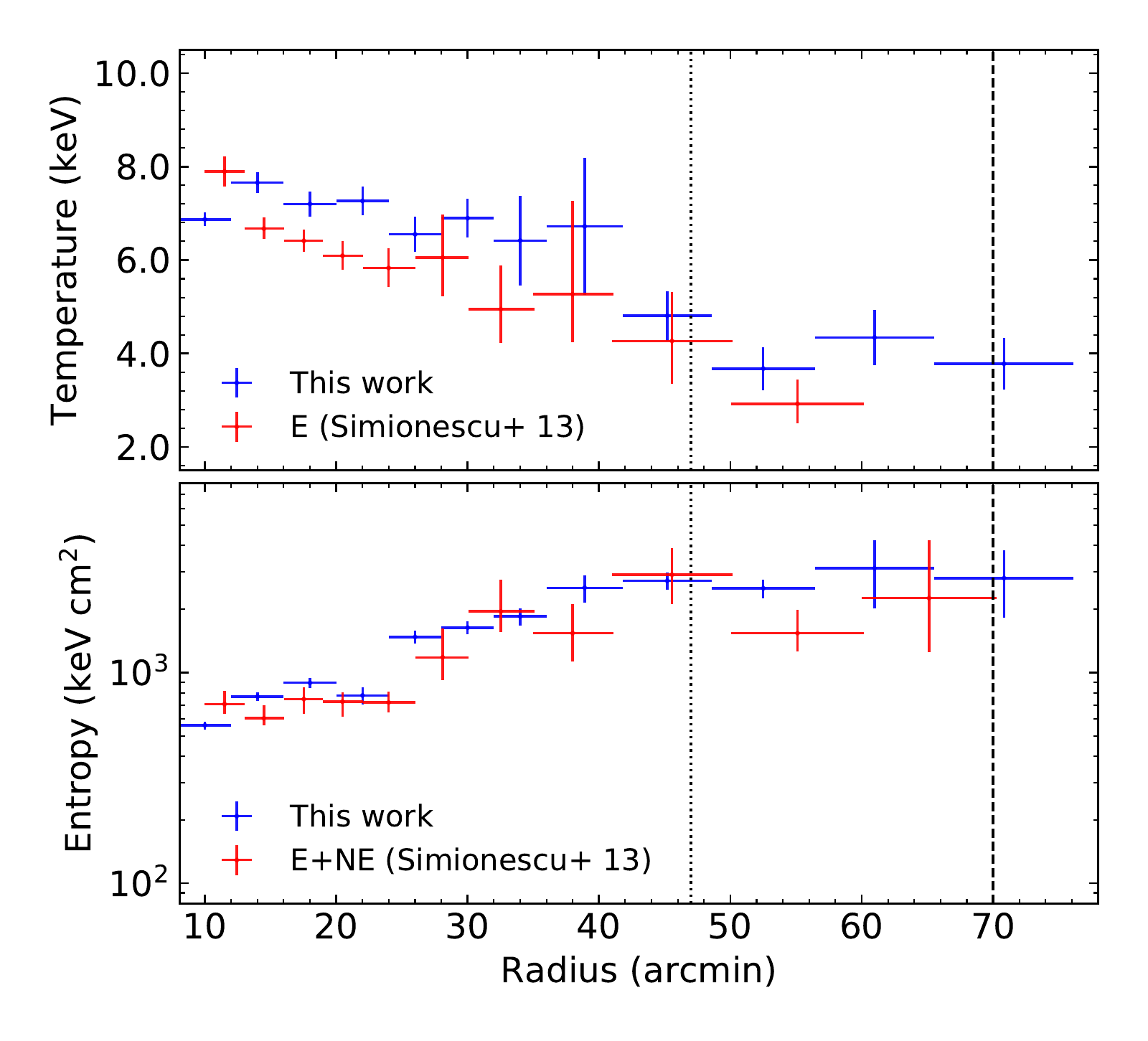}

\end{center}
\vspace{-0.5cm}
\caption{Comparing the Suzaku measurements of the temperature (\textit{top}) and entropy (\textit{bottom}) from \citet{simionescu2013thermodynamics} obtained along strips with the profiles along the same sectors found in this work. The error bars are the 1$\sigma$ percentiles computed using a Monte Carlo technique. The vertical dotted and dashed lines mark the positions of the $r_{500}$ and $r_{200}$ radii, respectively.  
  }
\label{fig: temp_entropy_comparison}
\vspace{-0.3cm}
\end{figure}

In Fig. \ref{fig: temp_entropy_comparison}, we compare our temperature and entropy measurements with those obtained using \textit{Suzaku}, reported by \citet{simionescu2013thermodynamics}. Our measurements are extracted in the same regions as the \textit{Suzaku} observations to allow a fair comparison of the measurements. We find that there is good agreement between our temperature and entropy measurements between the two different methods. Our slightly higher entropies  in the outskirts are to be expected as our gas densities have been corrected for gas clumping.

\section{Discussion}
\label{sec: discussion}

In this work, we combined the gas density measurements obtained from the deprojection of the surface brightness radial profile using \textit{XMM-Newton} observations with the gas pressure measurements obtained through the SZ effect signal using the \textit{Planck} observatory. The radial profiles of the thermodynamic properties of the Coma cluster were recovered out to the virial radius with nearly full azimuthal coverage for the first time. Furthermore, our high-quality data allowed us to investigate in detail the thermodynamic properties of the ICM gas out to the Coma outskirts in the 36 azimuthal sectors. This has enabled us to produce the most complete high-resolution view of the thermodynamic properties of the outskirts of a galaxy cluster ever achieved.

\subsection{Complex morphology of Coma}

\begin{figure}
\begin{center}

\includegraphics[width=1.0\linewidth]{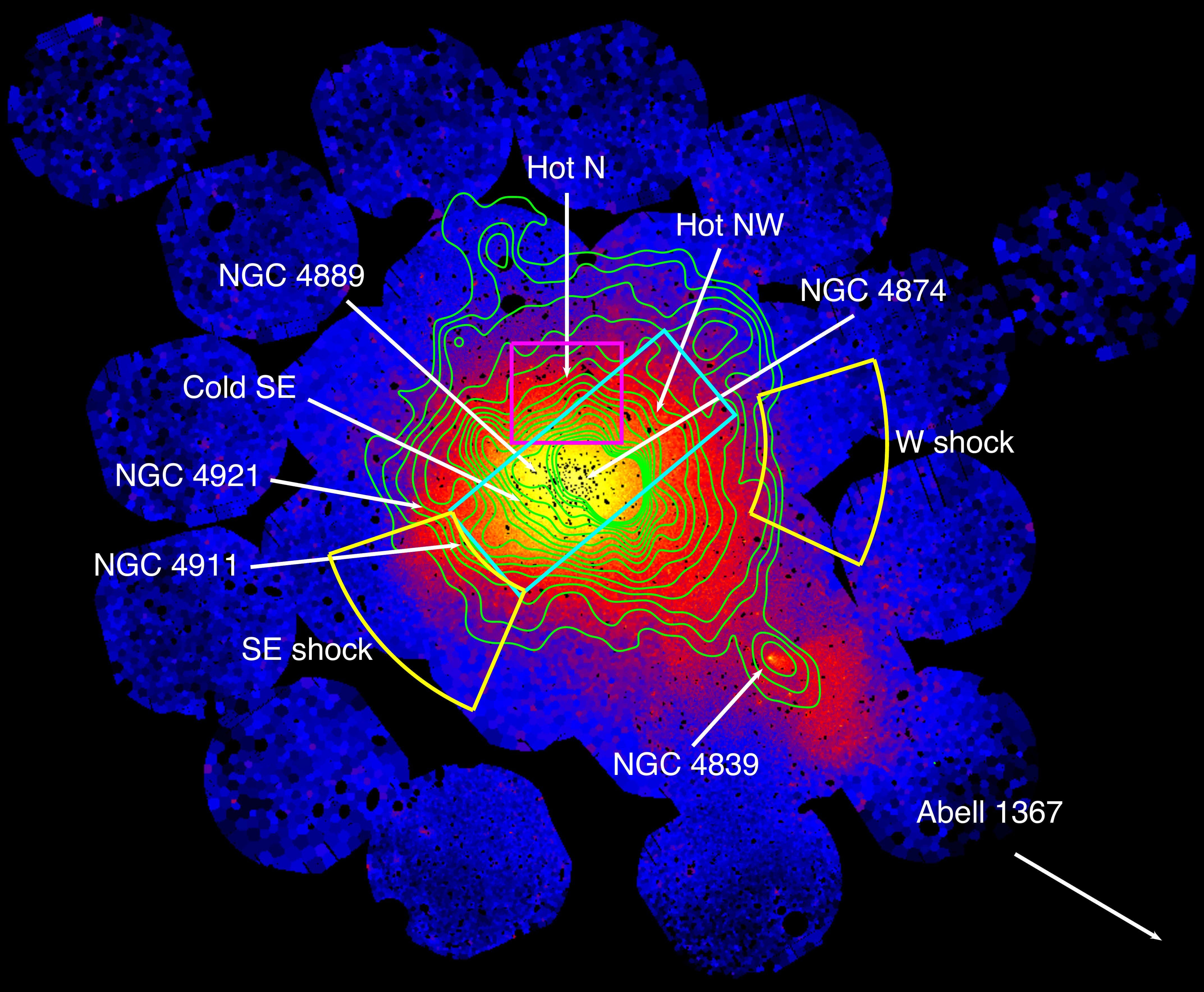}

\end{center}
\caption{\textit{XMM-Newton} image of the Coma cluster with the 352 MHz WSRT radio contours from \citet{brown2011diffuse} are overlaid. It shows the locations of the substructures and features of the cluster discussed in the main text.
  }
\label{fig: coma_xray_radio}
\vspace{-0.3cm}
\end{figure}

It is well established that the Coma cluster is an unrelaxed system, with a number of substructures which we summarize here and show in Fig. \ref{fig: coma_xray_radio}. 
\citet{lyskova2019close} have argued that the morphology of the infalling group NGC 4839 is best explained if it has fallen from the north east and already passed by the Coma centre, and is returning from the southwest direction. On the other hand, previous work (e.g.  \citealt{neumann2001ngc}, \citealt{Akamatsu2013}) has suggested that it is falling into Coma for the first time. 

Coma features two brightest cluster galaxies in its central regions, NGC 4874 and NGC 4889 \citep{Sanders2014}, which appear to be in their final stage of the merging process \citep{coccato2010kinematics}. \citet{vikhlinin1997another} have reported on a 1 Mpc long excess in X-ray surface brightness extending from the core to the south east direction towards the galaxy NGC 4911.

The Coma cluster hosts a giant radio halo \citep[Coma C;][]{willson1970radio} extending over scales of about 1 Mpc ($\approx$ 36 arcmin) and traces the non-thermal emission from relativistic electrons and magnetic fields (see Fig. \ref{fig: coma_xray_radio}). The Coma cluster also appears to have at least two shock fronts \citep{ade2013planck} in two regions about 40 arcmin to the west and the southeast of the cluster centre (marked in Fig. \ref{fig: coma_xray_radio}), corresponding to the outermost edge of the giant radio halo. 


\begin{figure}
\begin{center}
\vbox{
\includegraphics[width=1.0\linewidth]{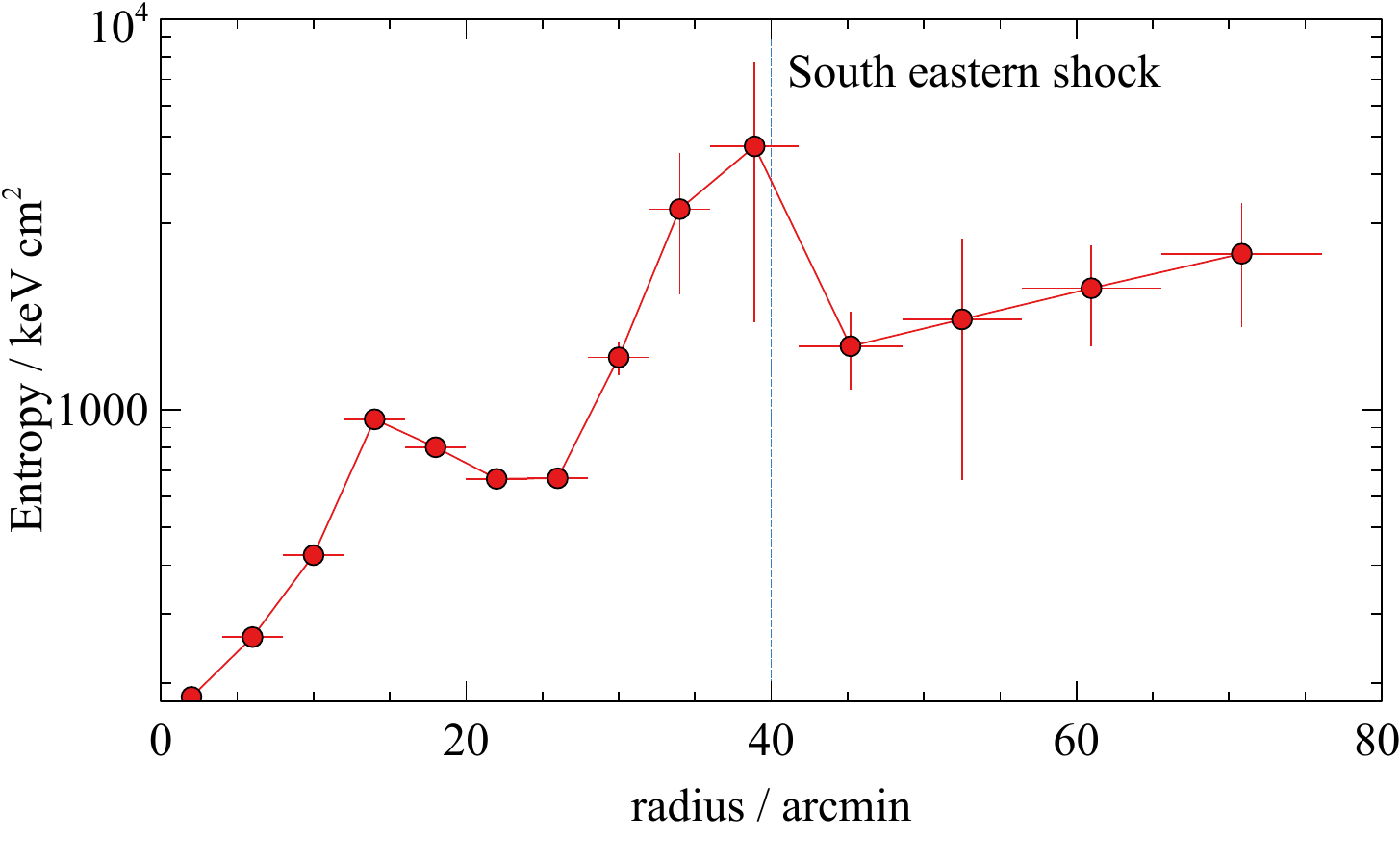}
\includegraphics[width=1.0\linewidth]{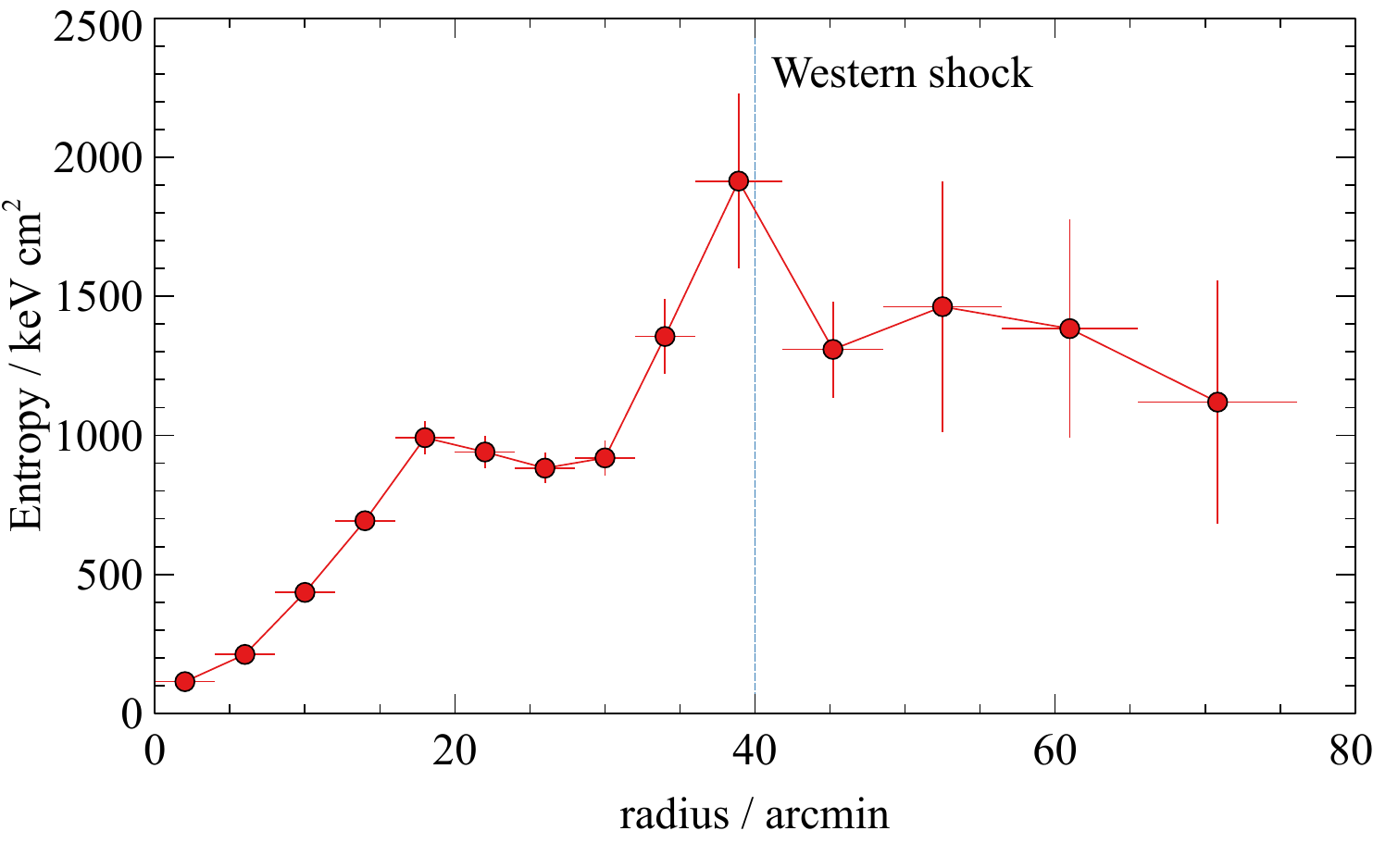}

}
\end{center}
\vspace{-0.5cm}
\caption{Entropy profiles along the directions of the South Eastern (top) and Western (bottom) shocks, showing the entropy jump behind the shock locations, which are marked by the vertical dashed line
  }
\label{fig: shock_profiles}
\vspace{-0.3cm}
\end{figure}

\subsubsection{Structure outside $r_{500}$}

In the south western direction, we have found that there is a clear entropy deficit (in the range 2-3$\sigma$) below the baseline entropy profile from 0.5$r_{200}$ outwards, and extending from sectors 27 to 36 (a quarter of the cluster azimuth) as shown in the entropy profiles in Fig. \ref{fig: entropy_sectors} and the entropy map in Fig. \ref{fig: temp_entropy_map2}. The extent of this entropy decrement is far larger than the size of the infalling group NGC 4839, which we exclude from our analysis.

This direction coincides with a where a cosmic web filament joins the Coma cluster, connecting it to Abell 1367 to the west \citep{malavasi2020like}. The locations of the filament galaxies surrounding the Coma cluster found in \cite{Mahajan2018} are shown as the green dots in Fig. \ref{fig: temp_entropy_map2}. Fig. \ref{fig: temp_entropy_map2} shows that the width and location of the galaxy distribution in the western filament is consistent with the size of the low entropy south western sector in Coma. This low entropy is consistent with what is expected for a penetrating filamentary gas stream \citep{zinger2016role}, which can bring low entropy gas deep into central regions of the cluster.

\citet{malavasi2020like} also report on two other cosmic web filaments connecting to the Coma cluster to the north and north east (shown by the green arrows and the galaxy distribution in Fig. \ref{fig: temp_entropy_map2}), however we find no entropy decrement in those directions. As described in \citet{zinger2016role}, the penetration of the gas streams into the cluster interior as the cluster ages can vary depending on a range of factors such as the mass inflow rate along the filament. In some cases filaments that once penetrated into the cluster core can weaken and be `pushed out’ to radii greater than the virial radius. It is therefore possible that the south western filamentary gas stream has been able to penetrate deeper into the cluster than the other two filaments, possibly due to the mass inflow rate being higher along the south western filament. 

\citet{zinger2016role} also find in their simulations that deeply penetrating gas streams can carry large amounts of momentum and energy into the cluster core, driving random motions and keeping the cluster in an unrelaxed dynamical state. It is possible that the deeply penetrating accretion along the south western filament in the Coma cluster is at least partly responsible for keeping it in an unrelaxed (non cool core) state.

\subsubsection{Inside $r_{500}$}

Our measurements within a radius of 30 arcmin from the cluster centre, on average, show a temperature gradient from the hotter northwest region to the colder region in the southeast, in agreement with previous studies \citep[e.g.][]{watanabe1999temperature,neumann2003dynamical,sato2011suzaku,sanders2013linear}. In Fig. \ref{fig: temp_entropy_map}, we show the temperature map of the central region of Coma from our joint \textit{XMM/Planck} analysis, and compare it to the \textit{Chandra} temperature map of the core from \citet{sanders2013linear}. The asymmetry between the hotter northern region and cooler southern region found in the core with \textit{Chandra} is found to extend out to much larger radius. 

\citet{watanabe1999temperature} and \citet{neumann2003dynamical} argued that the hot region in the northwest is due to shock waves induced by the infall of a small substructure around NGC 4874 to the main body of the Coma cluster from the southeast direction. On the other hand, the cool gas in the southeast region, particularly near 25 arcmin, shows a significant drop in the entropy profile, relative to the averaged profile, which seems to be mainly caused by the enhanced gas density in the region. \citet{sanders2013linear} argue that the cool gas in this region is likely stripped from the infall of a small group, consisting of low-entropy material, centered near the two galaxies NGC 4921 and NGC 4911.  

\citet{donnelly1999hot} and \citet{neumann2003dynamical} observed a hot region within a radius of 30 arcmin to the north from the centre of the Coma cluster (Fig. \ref{fig: temp_sectors}, upper panels) using \textit{ASCA} and \textit{XMM} respectively. Our results confirm the existence of this hot spot, as reported by these authors. In the same region, we also observe an apparent entropy excess (Fig. \ref{fig: entropy_sectors}, upper panels) and a low gas mass fraction compared with other regions (Fig. \ref{fig: gas_mass_fraction_sectors}, upper panels). It was argued that this region is heated by a shock wave created by a previous merger from a small group centered near the galaxy NGC 4889, which is currently one of the two dominant galaxies in the core of the Coma cluster. 

As expected from the shocks observed by \citet{ade2013planck} to the south-east and west, at radii between 30-40 arcmin, the southeastern sectors show a jump in the gas temperature (Fig. \ref{fig: temp_sectors}, lower left) and entropy at the 1$\sigma$ level (Fig. \ref{fig: entropy_sectors}, lower left and also Fig. \ref{fig: shock_profiles} in which we show just the entropy profiles in the directions of the shocks). At the same radii, the radial profile of the gas density along these sectors shows a sharp drop (Fig. \ref{fig: ne_sectors}, lower left). A similar trend is also seen in the thermodynamic properties at the same radii along the western sectors. The structure at 10-15 arcmin to the SE is caused by a linear feature near the core that has been studied in depth in previous works (e.g. \citealt{sanders2013linear}).

\section{Conclusions}
\label{sec: conclusions}

In this work, we have presented our new extended \textit{XMM-Newton} mosaic of the Coma cluster, which covers the cluster out to $r_{200}$ with nearly complete azimuthal coverage. By combining the gas density measurements obtained from the X-ray surface brightness profile of this mosaic with the gas pressure measurements obtained via the SZ effect signal using \textit{Planck}, we have able to recover the thermodynamic properties of the Coma cluster out to the virial radius with nearly full azimuthal coverage for the first time. Furthermore, our high-quality X-ray and SZ effect data have allowed us to investigate in detail the thermodynamic properties of the ICM out to $r_{200}$ in 36 azimuthal sectors. This has enabled us to produce the highest spatial resolution view of the thermodynamics of the entire outskirts of a galaxy cluster ever achieved.

Beyond $r_{500}$, the temperature profiles of the Coma cluster, on average, do not appear to drop sharply. The entropy profiles along the relatively less disturbed directions are statistically consistent with the baseline entropy profile predicted by cosmological simulations for purely gravitational hierarchical structure formation \citep{Voit2005}.

Our results show that the radial profiles of the thermodynamic quantities recovered from X-ray and SZ effect data become increasingly asymmetric in the outskirts, in agreement with the predictions of cosmological simulations of galaxy cluster formation \citep[e.g.][]{avestruz2014testing,lau2015mass}.

We find that there is a significant entropy deficit from 0.5$r_{200}$ out to the virial radius in the south western direction, which coincides with the direction of a cosmic web filament which connects the Coma cluster to Abell 1367 (Fig. \ref{fig: temp_entropy_map2}). The entropy deficit extends for around a quarter of the cluster azimuth to the south west. The size of the entropy decrement is very similar to the width of the western filament as traced by its constituent galaxies (around 2 Mpc), and is much larger than the spatial scale of the infalling group NGC 4839 which we exclude from the analysis. This low entropy is consistent with theoretical expectations for filamentary gas streams originating from the cosmic web (e.g. \citealt{zinger2016role}), which can penetrate deep into the cluster interior, channeling low entropy gas. 

The inner radius to which this entropy decrement extends is consistent with the radius of the western shock, 40 arcmin from the core. As remarked upon by \cite{malavasi2020like}, who studied the filamentary large scale environment of Coma and observed that the western shock location coincides with the connection between Coma and the western filament, it is possible that this western shock is the result of accretion along the western cosmic web filament onto the Coma cluster. 

The two other cosmic web filaments connecting to the Coma cluster to the north and northeast do not appear to have significantly disrupted the thermodynamics of the Coma cluster within the virial radius. It is possible that the mass inflow rate is much lower along those filaments, and that their gas streams are currently unable to penetrate into the central $r_{200}$ of the cluster. Simulations \citep{zinger2016role} have found the penetration of the gas streams associated with cosmic web filaments can vary substantially, with some penetrating deep into the central regions of clusters while other are unable to penetrate within the virial radius. If we are seeing a gas stream penetrating the Coma cluster to the south west, it is possible that the large amounts of energy and momentum it carries into the cluster is contributing to keeping Coma in an unrelaxed state.

At small and intermediate radii, our results confirm the existence of several complex features reported in the literature. The measurements within a radius of 30 arcmin show a gradient in the gas temperature from the hotter northwest region to the colder region in the southeast, in agreement with previous studies \citep[e.g.][]{watanabe1999temperature,neumann2003dynamical}. These high temperature regions may be due to shock heating induced by the infall of a substructure around NGC 4874 to the cluster core from the southeast direction. Our observations also confirm the existence of a hot spot at the north side of the Coma core, which reported in previous studies using \textit{ASCA} \citep[e.g.][]{donnelly1999hot}.

 At radii between 30-40 arcmin, there is a jump in the gas temperature and entropy along the southeastern sectors.  A similar trend is also seen in the thermodynamics at the same radii along the western sectors. These features correspond to the well known shock fronts \citep{ade2013planck}, and the edge of these features coincides with the edge of the giant radio halo observed at 352 MHz \citep{brown2011diffuse}.

Our results underline the enormous discovery space presented by the outskirts of galaxy clusters to new high throughput X-ray observatories such as \textit{ATHENA} (\citealt{ATHENA}), and proposed concepts such as \textit{Lynx} (\citealt{Lynx}) and \textit{AXIS} (\citealt{AXIS}). A complete understanding of how galaxy clusters form and grow can only be achieved by pushing out to understand their outskirts and how they connect to the surrounding cosmic web (\citealt{WalkerWhitePaper}, \citealt{BulbulWhitePaper}, \citealt{SimionescuWhitePaper}).

\section*{Acknowledgements}
We thank the referee for their helpful report. MSM and SAW acknowledge support from the NASA XMM-Newton grant 19-XMMNC18-0030. Based on observations obtained with XMM-Newton, an ESA science mission with instruments and contributions directly funded by ESA Member States and NASA. We thank Shea Brown for providing the radio contours shown in Fig. \ref{fig: coma_xray_radio}, and Jeremy Sanders for providing the central \textit{Chandra} temperature map shown in Fig. \ref{fig: temp_entropy_map}. 

\section*{Data Availability}

The XMM-Newton Science Archive (XSA) stores the archival data used in this paper, from which the data are publicly available for download.  The XMM data were processed using the XMM-Newton Science Analysis System (SAS). The
software packages heasoft and xspec were used, and these can be downloaded from the High Energy Astrophysics Science Archive
Research Centre (HEASARC) software web-page. Analysis and figures were produced using \textsc{python} version 3.7.

\bibliographystyle{mn2e}
\bibliography{Coma}

\appendix

\section{\textit{XMM-Newton} observations of Coma}
Table \ref{tab: xmm_observations} shows the details of all \textit{XMM-Newton} observations used in this work. The outskirts and background pointings are labeled with the words "Outskirts" and "Background", respectively, followed by numerical indices that increase counterclockwise from the Right Ascension (RA) axis. The remaining pointings that cover the central and intermediate regions up to $r_{500}$ are labeled with the word "Inside", followed by numerical indices. In this table, all outskirts and background observations are new observations, except the observations that labeled "Outskirts10", "Outskirts11", and "Outskirts12". The outskirts pointings that has the same RA and Dec. coordinate set to have the same numerical indices.  

\begin{table*}
    \centering
    \caption{\textit{XMM-Newton} observations of Coma}
    \begin{tabular}{lccccc}
    \hline
    
   Observation & Obs. ID & Obs. Date & RA & Dec. & Exposure \\
     &   &  & (J2000) & (J2000)   & (ks)  \\
    \hline
    Outskirts1    & 0841680101 & 07 Jul 2019 & 12 55 25.59 & +27 47 01.6 & 23.0 \\ 
    Outskirts2    & 0841680201 & 07 Jul 2019 & 12 55 38.07 & +28 17 43.1 & 23.0 \\
    Background1   & 0841681101 & 12 Jul 2019 & 12 53 31.92 & +28 29 54.4 & 29.5 \\
    Outskirts3    & 0841680301 & 11 Jul 2019 & 12 57 03.20 & +28 41 26.5 & 26.0 \\
    Outskirts4    & 0841680401 & 13 Jul 2019 & 12 59 11.76 & +28 52 56.4 & 23.0 \\
    Outskirts5    & 0841680501 & 13 Jul 2019 & 13 01 27.90 & +28 51 36.7 & 23.0 \\
    Outskirts6    & 0841680601 & 17 Jul 2019 & 13 03 15.79 & +28 32 54.7 & 23.0 \\
    Background2   & 0841681201 & 29 Dec 2019 & 13 05 10.70 & +28 53 08.8 & 23.0 \\
    Outskirts7    & 0841680701 & 16 Jul 2019 & 13 04 03.83 & +28 04 45.1 & 26.4 \\
    Outskirts8    & 0841680801 & 17 Jul 2019 & 13 03 56.84 & +27 35 23.5 & 34.4 \\
    Outskirts9    & 0841680901 & 06 Dec 2019 & 13 02 49.88 & +27 11 18.6 & 30.8 \\
    Outskirts10   & 0800580101 & 23 Dec 2017 & 13 00 57.97 & +26 58 35.3 & 80.4 \\
    Outskirts10   & 0800580201 & 04 Jan 2018 & 13 00 57.97 & +26 58 35.3 & 87.0 \\
    Outskirts11   & 0403150301 & 17 Jun 2006 & 12 57 40.75 & +26 56 13.6 & 56.0 \\
    Outskirts11   & 0403150401 & 21 Jun 2006 & 12 57 40.75 & +26 56 13.6 & 64.4 \\
    Outskirts12   & 0058940701 & 10 Jun 2003 & 12 55 24.99 & +27 13 46.0 & 22.6 \\
    Inside1         & 0124710501 & 29 May 2000 & 12 59 27.49 & +27 46 53.0 & 29.9 \\
    Inside2        & 0124711401 & 29 May 2000 & 12 59 46.71 & +27 56 60.0 & 34.6 \\
    Inside3         & 0124711601 & 11 Jun 2000 & 12 57 42.51 & +27 43 38.0 & 87.9 \\
    Inside4         & 0124710201 & 11 Jun 2000 & 12 57 42.51 & +27 43 38.0 & 41.5 \\
    Inside5         & 0124710901 & 11 Jun 2000 & 13 00 32.68 & +27 56 59.0 & 31.2 \\
    Inside6         & 0124710601 & 12 Jun 2000 & 12 58 50.01 & +27 58 52.0 & 31.8 \\
    Inside7         & 0124710101 & 21 Jun 2000 & 12 56 47.68 & +27 24 07.0 & 41.5 \\
    Inside8         & 0124710401 & 23 Jun 2000 & 13 00 04.60 & +27 31 24.0 & 52.8 \\
    Inside9         & 0124711101 & 24 Jun 2000 & 12 58 36.51 & +28 23 56.0 & 40.0 \\
    Inside10         & 0124710701 & 24 Jun 2000 & 12 57 27.68 & +28 08 41.0 & 27.2 \\
    Inside11         & 0124710301 & 27 Jun 2000 & 12 58 32.19 & +27 24 12.0 & 28.6 \\
    Inside12         & 0124712201 & 09 Dec 2000 & 12 57 42.51 & +27 43 38.0 & 27.6 \\
    Inside13         & 0124712001 & 10 Dec 2000 & 12 58 50.01 & +27 58 52.0 & 22.8 \\
    Inside14         & 0124712101 & 10 Dec 2000 & 12 57 27.68 & +28 08 41.0 & 28.1 \\
    Inside15         & 0124710801 & 10 Dec 2000 & 13 01 25.60 & +27 43 53.0 & 29.8 \\
    Inside16         & 0153750101 & 04 Dec 2001 & 12 59 46.71 & +27 56 60.0 & 25.8 \\
    Inside17         & 0124712401 & 05 Jun 2002 & 13 01 50.19 & +28 09 28.0 & 27.6 \\
    Inside18         & 0124712501 & 07 Jun 2002 & 13 00 36.50 & +28 25 15.0 & 28.7 \\
    Inside19         & 0204040101 & 06 Jun 2004 & 13 00 22.21 & +28 24 03.0 & 101.9 \\
    Inside20         & 0204040201 & 18 Jun 2004 & 13 00 22.21 & +28 24 03.0 & 108.3 \\
    Inside21         & 0204040301 & 12 Jul 2004 & 13 00 22.21 & +28 24 03.0 & 104.2 \\
    Inside22         & 0300530701 & 06 Jun 2005 & 12 59 36.92 & +27 58 14.8 & 25.5 \\
    Inside23         & 0300530601 & 07 Jun 2005 & 12 59 35.41 & +27 56 33.3 & 25.7 \\
    Inside24         & 0300530501 & 08 Jun 2005 & 12 59 39.73 & +27 55 12.0 & 25.5 \\
    Inside25         & 0300530401 & 09 Jun 2005 & 12 59 46.67 & +27 55 12.0 & 27.5 \\
    Inside26         & 0300530301 & 11 Jun 2005 & 12 59 51.00 & +27 56 33.3 & 31.0 \\
    Inside27         & 0300530201 & 17 Jun 2005 & 12 59 49.45 & +27 58 14.8 & 27.5 \\
    Inside28         & 0300530101 & 18 Jun 2005 & 12 59 43.18 & +27 58 59.8 & 25.5 \\
    Inside29         & 0304320301 & 27 Jun 2005 & 13 00 22.21 & +28 24 03.0 & 55.9 \\
    Inside30         & 0304320201 & 28 Jun 2005 & 13 00 22.21 & +28 24 03.0 & 80.8 \\
    Inside31         & 0304320801 & 06 Jun 2006 & 13 00 22.21 & +28 24 03.0 & 63.8 \\
    Inside32         & 0403150201 & 11 Jun 2006 & 12 57 42.51 & +27 19 09.7 & 55.2 \\
    Inside33         & 0403150101 & 14 Jun 2006 & 12 57 42.51 & +27 19 09.7 & 54.4 \\
    Inside34         & 0652310701 & 16 Jun 2010 & 12 57 24.29 & +27 29 52.0 & 21.8 \\
    Inside35         & 0652310201 & 18 Jun 2010 & 12 57 24.29 & +27 29 52.0 & 22.3 \\
    Inside36         & 0652310301 & 20 Jun 2010 & 12 57 24.29 & +27 29 52.0 & 19.4 \\
    Inside37         & 0652310401 & 24 Jun 2010 & 12 57 24.29 & +27 29 52.0 & 23.9 \\
    Inside38         & 0652310501 & 04 Jul 2010 & 12 57 24.29 & +27 29 52.0 & 22.9 \\
    Inside39         & 0652310601 & 06 Jul 2010 & 12 57 24.29 & +27 29 52.0 & 20.8 \\
    Inside40         & 0652310801 & 03 Dec 2010 & 12 57 24.29 & +27 29 52.0 & 16.9 \\
    Inside41         & 0652310901 & 05 Dec 2010 & 12 57 24.29 & +27 29 52.0 & 16.9 \\
    Inside42         & 0652311001 & 11 Dec 2010 & 12 57 24.29 & +27 29 52.0 & 16.4 \\
    Inside43         & 0691610201 & 02 Jun 2012 & 12 57 24.65 & +27 29 42.7 & 37.9 \\
    Inside44         & 0691610301 & 04 Jun 2012 & 12 57 24.65 & +27 29 42.7 & 35.9 \\
    Inside45         & 0851180501 & 30 May 2019 & 13 02 00.14 & +27 46 57.8 & 48.4 \\
    \hline
    \end{tabular}
    \label{tab: xmm_observations}
\end{table*}

\section{Thermodynamic properties in azimuthal sectors}

\begin{figure*}
\begin{center}
\includegraphics[width=0.85\textwidth]{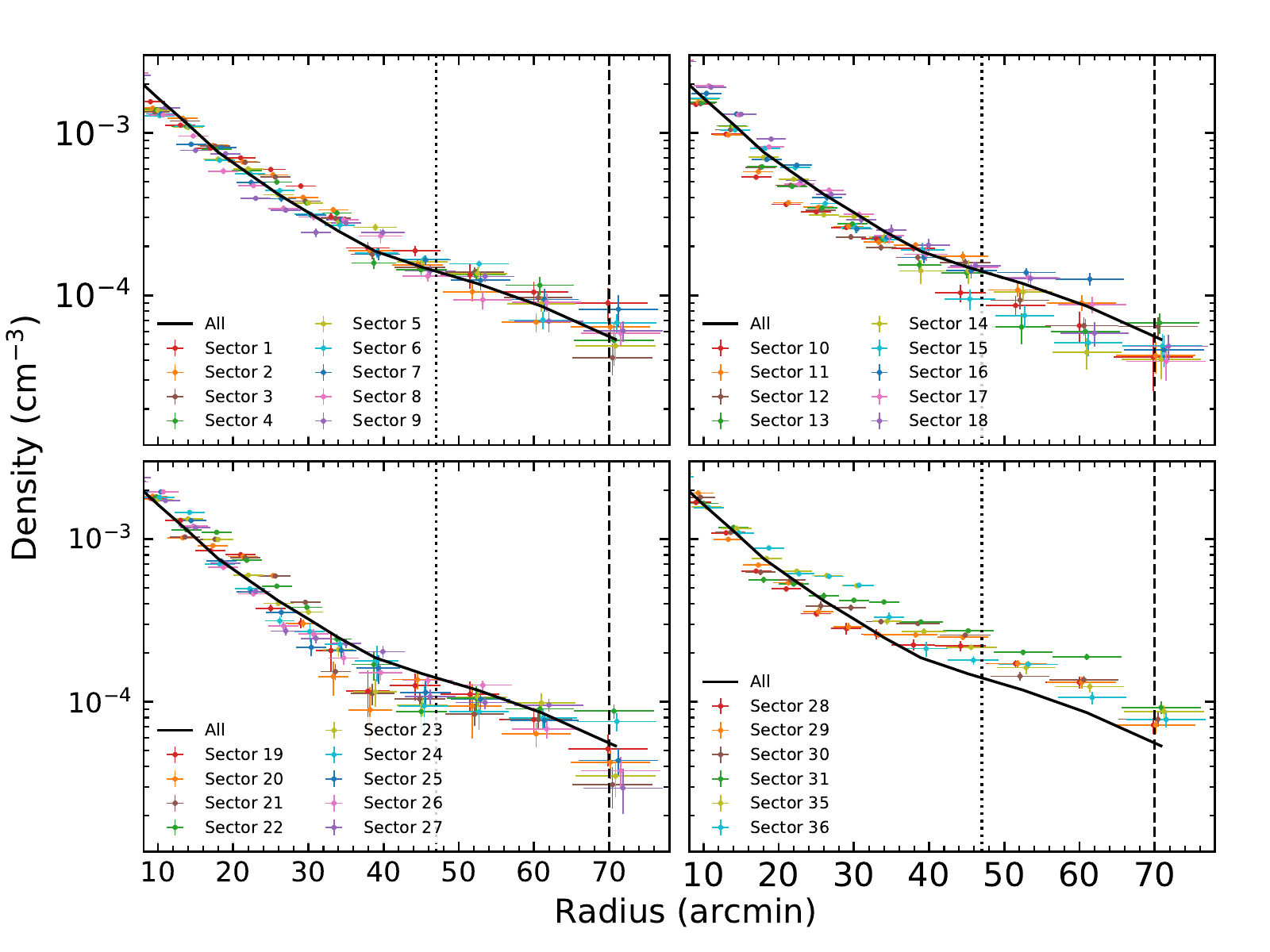}
\end{center}
\vspace{-0.5cm}
\caption{ Density profiles in the azimuthal sectors obtained from the deprojection of the surface brightness radial profiles using the onion peeling technique. The error bars are the 1 $\sigma$ percentiles computed using a Monte Carlo technique. The thick black line shows the azimuthally averaged density profile. The vertical dotted and dashed lines represent the $r_{500}$ and $r_{200}$ radii, respectively.
  }
\label{fig: ne_sectors}  
\vspace{-0.5cm}
\end{figure*}

\begin{figure*}
\begin{center}

\includegraphics[width=0.85\textwidth]{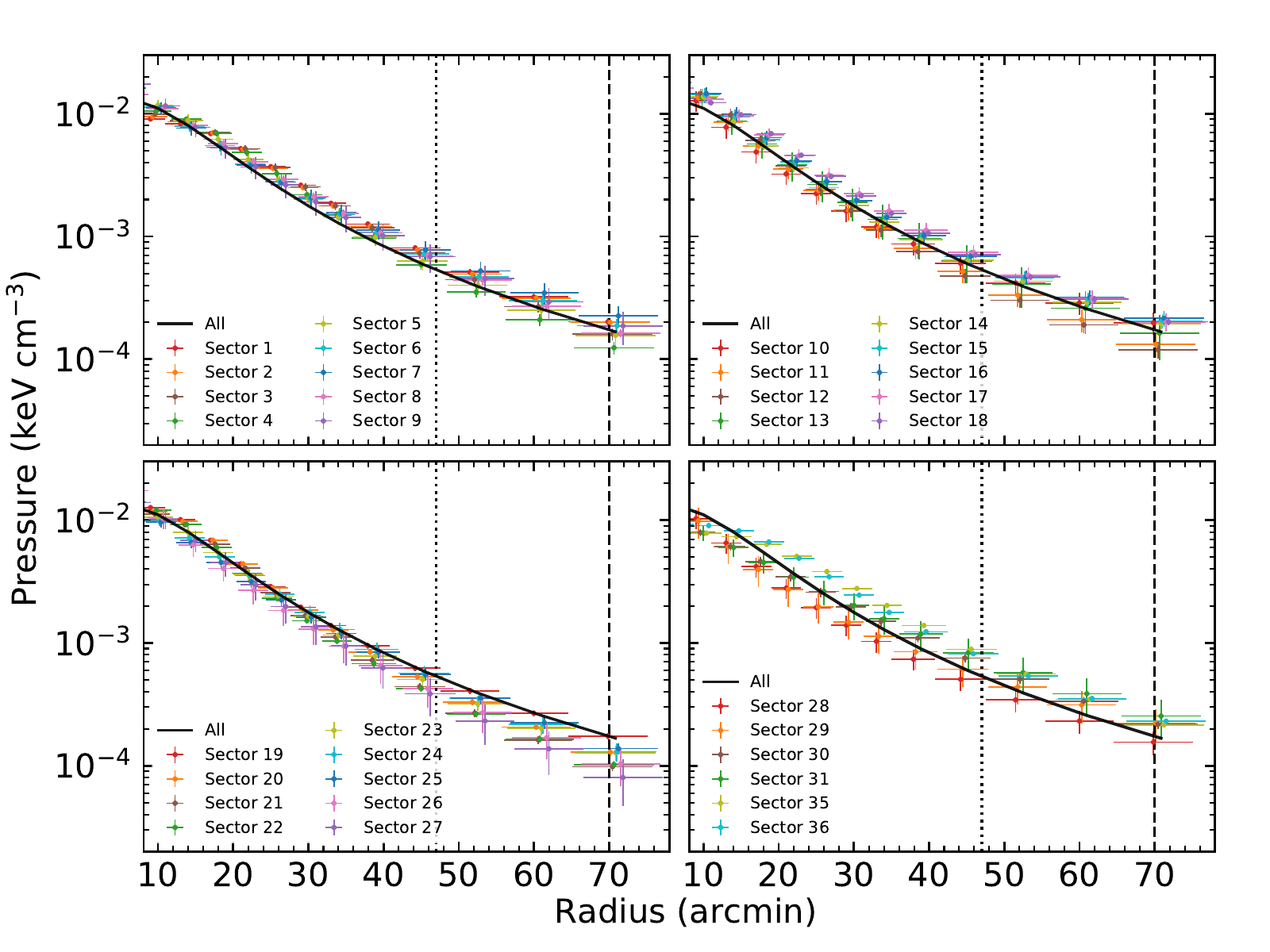}

\end{center}
\vspace{-0.5cm}
\caption{Same as Fig. \ref{fig: ne_sectors}, except for the gas pressure.
  }
\label{fig: pressure_sectors}
\vspace{-0.5cm}
\end{figure*}

\begin{figure*}
\begin{center}

\includegraphics[width=0.85\textwidth]{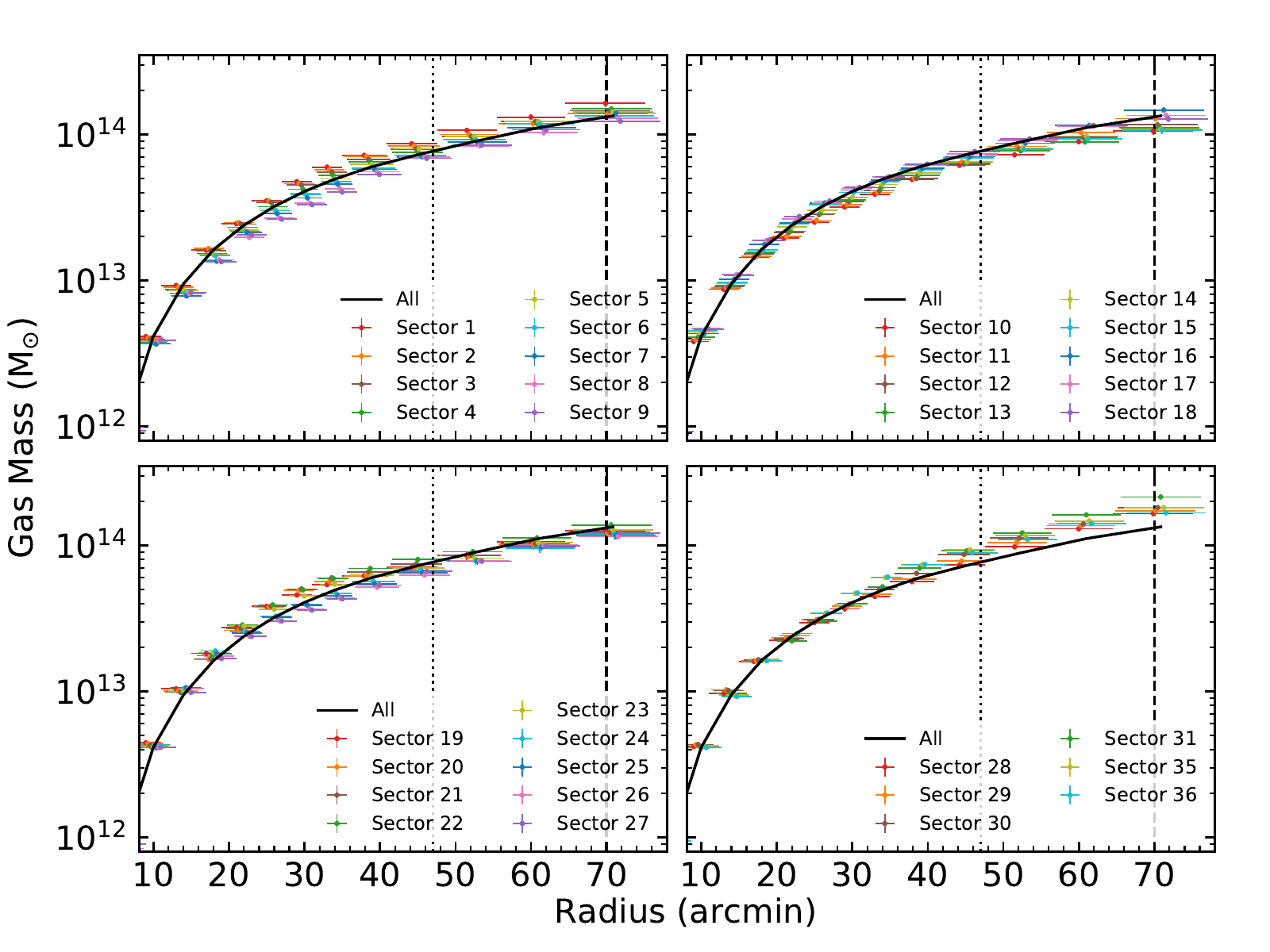}

\end{center}
\vspace{-0.5cm}
\caption{Same as Fig. \ref{fig: ne_sectors}, except for the gas mass.
  }
\label{fig: gas_mass_sectors}
\vspace{-0.5cm}
\end{figure*}

\begin{figure*}
\begin{center}
\includegraphics[width=0.85\textwidth]{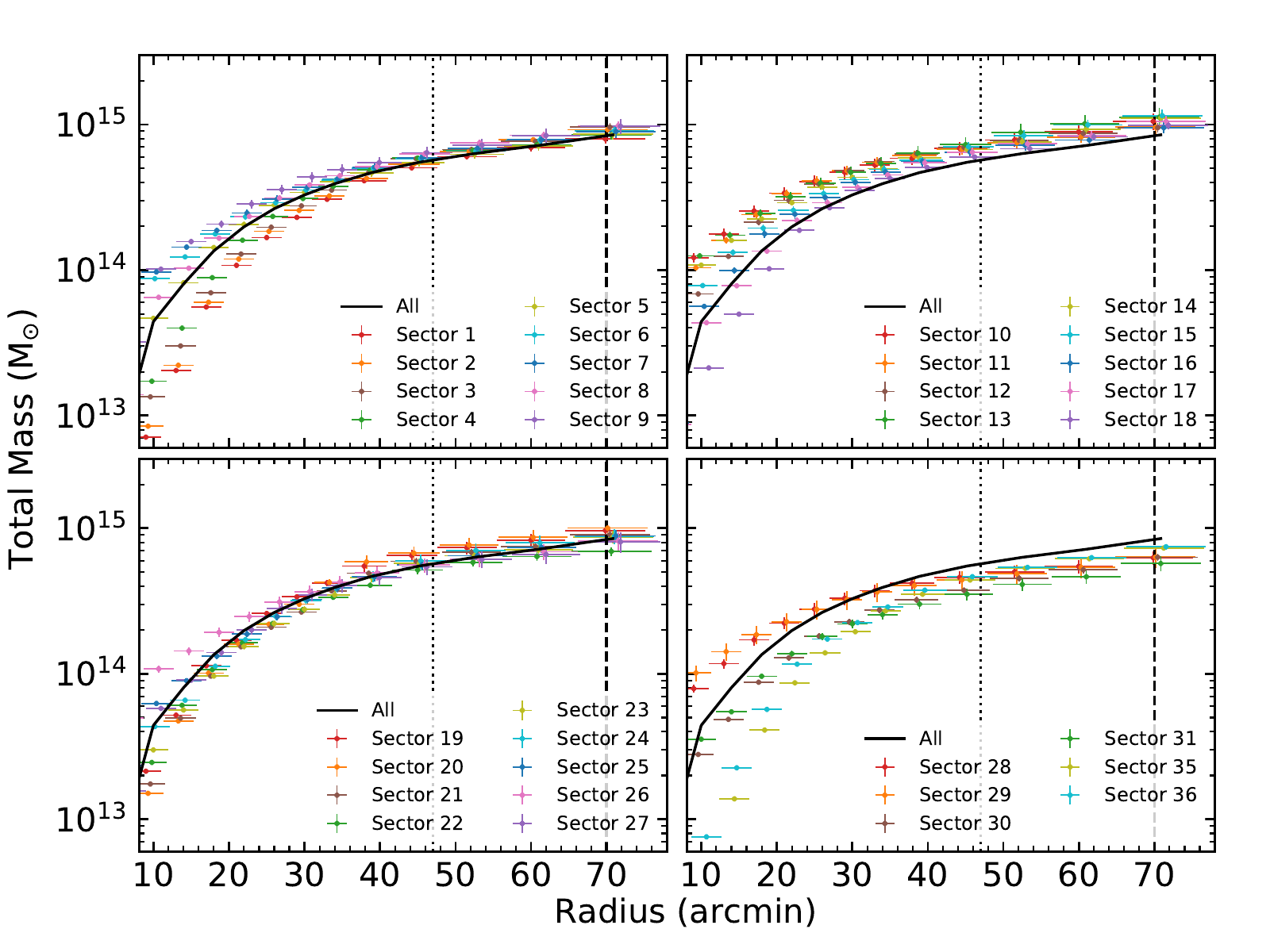}
\end{center}
\vspace{-0.5cm}
\caption{Same as Fig. \ref{fig: ne_sectors}, except for the total mass.
  }
\label{fig: tot_mass_sectors}
\vspace{-0.5cm}
\end{figure*}

In Fig. \ref{fig: ne_sectors}, we show the gas density profiles of the azimuthal sectors obtained from the deprojection of the radial profiles of the X-ray surface brightness using \textit{XMM-Newton} observations. The gas pressure profiles in the azimuthal sectors (Fig. \ref{fig: pressure_sectors}), on the other hand, obtained through the SZ effect signal using \textit{Planck} by fitting the projected $y$ profile in each sector to the universal pressure profile (equation \ref{equ: universal_pressure}). From the gas density and pressure profiles, as explained in Section \ref{sec: joint xray and sz}, the gas and total masses of the azimuthal sectors were recovered (see Figs \ref{fig: gas_mass_sectors} and \ref{fig: tot_mass_sectors}). All measurements reported in these figures take account of the systematic uncertainties.

\end{document}